\documentclass[12pt]{iopart}

\usepackage{iopams}  
\usepackage[breaklinks=true,colorlinks=true,linkcolor=blue,urlcolor=blue,citecolor=blue]{hyperref}
\usepackage{graphicx}
\usepackage{braket}
\usepackage{float}
\usepackage{caption}
\usepackage[]{algorithm2e}
\usepackage{enumitem}
\expandafter\let\csname equation*\endcsname\relax

\expandafter\let\csname endequation*\endcsname\relax

\usepackage{mathtools}
\usepackage{amsmath}
\usepackage[T1]{fontenc}
\usepackage{color, colortbl}
\usepackage[dvipsnames]{xcolor}
\usepackage[noadjust]{cite} 

\newcommand\GoodBitVectorgeneral{\chi}
\newcommand\GeneralSuccessProb{\alpha}

\newcommand\PHILTERtolerance{\tau}
\newcommand\stabilizer{s}
\newcommand\good{\mathcal{I}}
\newcommand\GoodSet{X_{d}}
\newcommand\GoodBitVector{d}
\newcommand\LenGoodBitVector{\text{len}(d)}

\newcommand\QPEerrorTolerance{\text{err}}
\newcommand\QPEbitstring{x}

\newcommand\AnsatzProbOverlap{a}
\newcommand\TrueProbOverlap{b}

\begin{document}

\title[]{Quantum Computation of Eigenvalues within Target Intervals}

\author{Phillip W. K. Jensen}
\address{Department of Chemistry, University of Toronto, Toronto, Ontario M5G 1Z8, Canada}
\ead{phillip.kastberg@gmail.com}

\author{Lasse Bj{\o}rn Kristensen}
\address{Department of Physics and Astronomy, Aarhus University, DK-8000 Aarhus C, Denmark}
\ead{lbk@phys.au.dk}

\author{Jakob S. Kottmann}
\address{Department of Chemistry, University of Toronto, Toronto, Ontario M5G 1Z8, Canada}
\ead{jakob.kottmann@gmail.com}

\author{Al\'{a}n Aspuru-Guzik}
\address{Department of Chemistry, University of Toronto, Toronto, Ontario M5G 1Z8, Canada}
\address{Department of Computer Science, University of Toronto, Toronto, Ontario M5G 1Z8, Canada}
\address{Vector Institute for Artificial Intelligence, Toronto, Ontario M5S 1M1, Canada}
\address{Canadian Institute for Advanced Research, Toronto, Ontario M5G 1Z8, Canada}
\ead{alan@aspuru.com}

\vspace{10pt}
\begin{indented}
\item[] 27 September 2020 
\end{indented}

\begin{abstract}
There is widespread interest in calculating the energy spectrum of a Hamiltonian, for example to analyze optical spectra and energy deposition by ions in materials. In this study, we propose a quantum algorithm that samples the set of energies within a target energy-interval without requiring good approximations of the target energy-eigenstates. We discuss the implementation of direct and iterative amplification protocols and give resource and runtime estimates. We illustrate initial applications by amplifying excited states on molecular Hydrogen.
\end{abstract}

%

\section{Introduction}

\begin{figure}[t] 
\centering  
\includegraphics[width=0.8\textwidth]{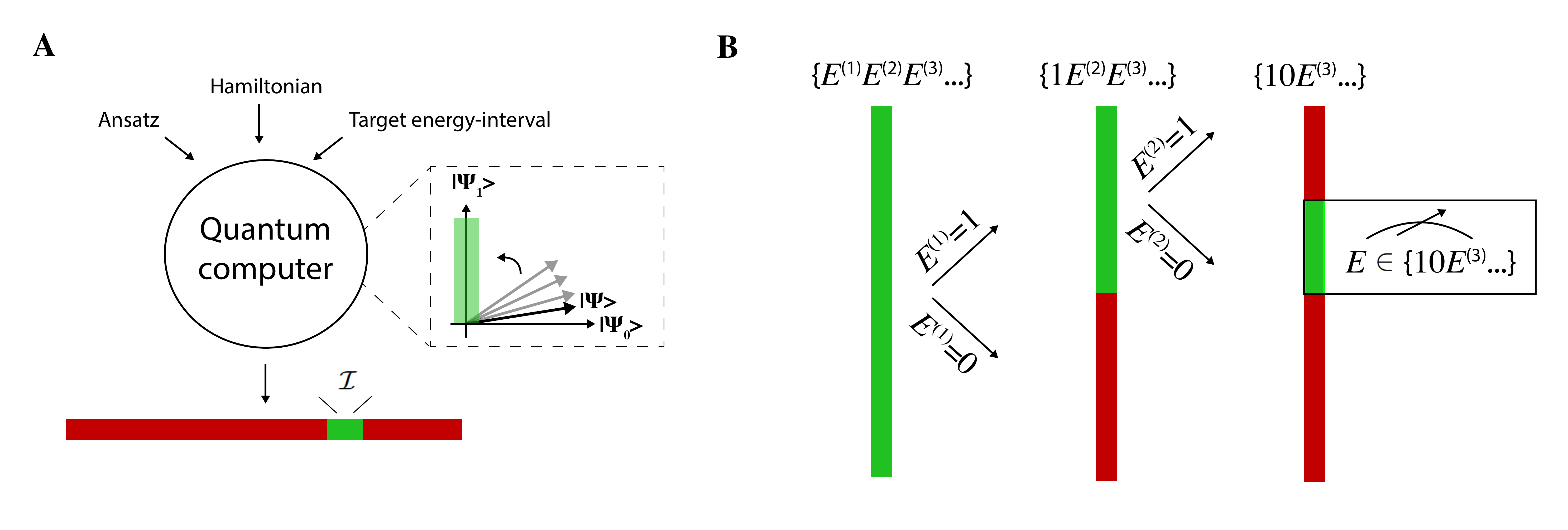} 
\caption{Schematic depiction of the algorithm. A: 
Given an Hamiltonian, an ansatz, target energy-interval and a quantum computer, our method drives the state $\ket{\Psi}$ to the target energy-interval $\good$. B: Schematic depiction of the operation of the algorithm in the case where an energy in the range $ \good = [0.5, 0.75) $ is desired, i.e. where the energy is denoted in binary and  the components $E^{(1)} = 1$ and $E^{(2)} = 0$ are picked out using amplitude amplification.}
\label{fig. SchematicRep}
\end{figure}

Quantum computers have the potential to leverage specific properties described by quantum mechanics to solve certain problems, such as prime factorization\cite{Shor1995}, search on unstructured data\cite{Grover1996, Grover1997} and decision trees\cite{Farhi1998, Farhi2007}, faster than known classical algorithms. One of the promising applications of quantum computing is to estimate energies for Hamiltonians, using for instance the quantum phase estimation (QPE) algorithm, originally proposed by Kitaev, Lloyd and Abrams\cite{Abrams1997, Abrams1999, Kitaev1997}, useful for chemical and physical problems. The molecular time-independent Schr\"{o}dinger equation provides an example of a fundamental eigenvalue problem suitable for QPE. Aspuru-Guzik \emph{et al.}\cite{AspuruGuzik2005} proposed the QPE method for solving the molecular time-independent Schr\"{o}dinger equation, and this approach was experimentally demonstrated by Lanyon \emph{et al.}\cite{Lanyon2010} using a photonic device to extract molecular properties using the iterative quantum phase estimation (IQPE) method\cite{Miroslav2007}. There is widespread interest in calculating the energy spectrum of a Hamiltonian, for example to understand optical spectra in quantum chemistry and energy deposition by ions in materials\cite{jensen2017}. Recently proposed methods for discovering Hamiltonian spectra make use of variational algorithms to find excited states\cite{Higgott2019, Jones2019, mcclean2016(1), lee2019, santagati2018}, for instance the witness-assisted variational eigenspectra solver (WAVES) protocol\cite{santagati2018} combines the variational method with phase estimation to find the excited states of Hamiltonians. For the WAVES method it is necessary to use an operator which approximates an excitation from the ground state to the desired excited state, an operator which may not be trivial to prepare. The orthogonally constrained variational quantum eigensolver with the unitary pair coupled cluster with generalized singles and doubles product ansatz is useful for obtaining low lying excited state energies\cite{ lee2019}.  However, the method has the drawback that it relies on the exact ground state or good approximations to it being known, and it solves for the excited state energies subsequently rendering it only feasible for low lying excited states. In this paper, we propose an algorithm that sample the set of energies for a given Hamiltonian within a target energy-interval without requiring good approximations of the target energy eigenstates. Using our approach, we drive an ansatz to a given energy-interval and measure the energies within the target energy-interval, as depicted in figure \ref{fig. SchematicRep}A. The ansatz is not restricted to approximate the target energy eigenstates, thus the algorithm is designed for cases where good approximations for the states in the target energy-interval are either unknown or hard to prepare. This would be useful for exploring excited state energies for Hamiltonians where it may not be possible to approximate the excited states of interest. Given a Hamiltonian, ansatz and target energy-interval, the algorithm amplifies the amplitudes for states in a restricted energy interval and reduces the probability for unwanted states. Our method is based on the studies on quantum amplitude amplification of Grover\cite{Grover1996, Grover1997}, Brassard \emph{et. al.}\cite{Brassard2000} and Boyer \emph{et al.}\cite{Boyer1998}.

The outline of the paper is as follows: First, in section \ref{sec:prelim} we give an overview of quantum phase estimation, quantum amplitude amplification and quantum amplitude estimation. In section \ref{sec: our_method} we present this paper's attendant ideas of combining these algorithms to compute energy eigenvalues within target energy-intervals. Finally in section \ref{sec. numerical}, as a proof-of-concept, an application to molecular Hydrogen is detailed, followed by a discussion and conclusion. 

\section{Preliminaries}
\label{sec:prelim}

\subsection{Quantum phase estimation}
\label{sec. QPE}

Assume we have been given a Hamiltonian $\mathcal{H}$ and would like to determine the energies within some given range. With a quantum computer we can use an operator $\emph{O}$ to prepare an ansatz that we believe to have some overlap with the kind of eigenstates we are looking for:

\begin{align}
\emph{O}\ket{0}^{\otimes n } = \sum_j a_j  \ket{E_j}, \label{Eq. ansatz}
\end{align}
where the state resulting from applying the operator has formally been written as a sum over the unknown energy eigenstates of $\mathcal{H}$ using coefficients $a_j$, and the operator acts on \emph{n} qubits  where \emph{n} depends on the encoding of the Hamiltonian. For example, for a molecular Hamiltonian we could prepare the Hartree-Fock (HF) state, $\emph{O}\ket{0}^{\otimes n } = \ket{\text{HF}}$, which is usually a starting point for more advanced methods. In order to extract information about the energies of the states, assume we add another \emph{m}-qubit register and run the QPE algorithm. The result of this will be that the \emph{m}-qubit register stores a binary representation of a phase related to the energies of the states 

\begin{align}
\ket{\Psi} &= \text{QPE}\big(\mathcal{H}\big)\emph{O}\ket{0}^{\otimes (m+n)} = \sum_j a_j \ket{E_j} \bigg(\sum^{2^m}_{\QPEbitstring_i \in \{0,1\}^m} \epsilon_{\QPEbitstring_i}^{(j)} \ket{\QPEbitstring_i} \bigg), \label{Eq. algorithm}
\end{align}
where for each energy eigenstate $\ket{E_j}$ we sum over bit strings $\QPEbitstring_i = \QPEbitstring_i^{(1)} \QPEbitstring_i^{(2)} \hdots \QPEbitstring_i^{(m)}$ with $\QPEbitstring_i^{(\alpha)} \in \{0,1\}$, and $\epsilon_{\QPEbitstring_i}^{(j)}$ being a set of complex coefficients. Let $E_j =E_j^{(1)} E_j^{(2)} E_j^{(3)} \hdots$ be the binary representation of the energy eigenvalue up to a rescaling factor.  The probability of observing the computational basis state $\ket{\QPEbitstring_i}$ in the \emph{m}-qubit energy register is then the expectation value of $\ket{\QPEbitstring_i} \bra{\QPEbitstring_i}$, which fulfills the inequality

\begin{equation}\label{eq. qpe_error}
\sum_{\{\QPEbitstring_i: \hspace{0.1cm}|E_j - \QPEbitstring_i|\leq  \QPEerrorTolerance  \}}\bra{\Psi}(I \otimes \ket{\QPEbitstring_i} \bra{\QPEbitstring_i})\ket{\Psi} \geq
\begin{cases} 
    |a_j|^2 \frac{8}{\pi^2} & \text{for $ \QPEerrorTolerance = 1$} \\
   |a_j|^2(1- \frac{1}{2(\QPEerrorTolerance-1)}) & \text{for $   \QPEerrorTolerance >1 $ }
   \end{cases} 
\end{equation} 
where $\QPEerrorTolerance \in \mathbb{Z}^+ $, $\QPEbitstring_i$ is an integer  $\QPEbitstring_i \in \{0,\hdots, 2^m-1\}$, and $ |E_j -  \QPEbitstring_i| $ is the difference between the energy eigenvalue $E_j$ and the measured value $\QPEbitstring_i$. That is, the probability that a measurement yields the best \emph{m}-bit approximation, i.e. within an accuracy of 1 to the true energy, is at least $ |a_j|^2 \frac{8}{\pi^2}$, where $a_j$ is the overlap amplitude between $\emph{O}\ket{0}^{\otimes n}$ and $\ket{E_j}$. The probability to obtain an energy output with error higher than 1 decreases as $\frac{1}{2(\QPEerrorTolerance-1)}$, which make QPE tolerant against errors because bit errors are more likely on the least significant bits in the binary representation of the energy. In the special case where the energy eigenvalues can be written exact with \emph{m}-bits then $\epsilon_{\QPEbitstring_i}^{(j)}=\delta_{x_i,E_j}$,  allowing only the correct bitstring to be measured. Throughout the paper, we refer the amplitudes $\epsilon_{\QPEbitstring_i}^{(j)}$ as the QPE amplitudes. A subroutine in the QPE is Hamiltonian simulation. Though Hamiltonian simulation is encoded exactly in this paper (section \ref{sec. numerical}), in general, an approximation must be made into a finite sequence of quantum gates\cite{Yudong2019}. The standard approaches are the Trotter and higher order Suzuki decompositions\cite{suzuki1990, suzuki1991,  berry2007}. We refer the reader to appendix C in \cite{cleve1998} for a detailed analysis of the success probability when estimating phases using the QPE. Thus what is stored in the \emph{m}-qubit register is essentially a binary representation of the energy up to a rescaling factor. What we would like to do to this state is to amplify the part of it that consists of states with energies within the target energy-interval and reduce all the components that do not have the right energies.

\subsection{Quantum amplitude amplification}
\label{sec. aa}
Assume the problem of interest is given by an operator  $\mathcal{A}$ acting on \emph{N} qubits such that

\begin{align}
 \ket{\Phi} = \mathcal{A}\ket{0}^{\otimes N} = \sqrt[]{1-\GeneralSuccessProb} \ket{\Phi_0} + \sqrt{\GeneralSuccessProb} \ket{\Phi_1} \label{Eq.general_QAAstate}
\end{align}
where 

\begin{align}
\ket{\Phi_0}  = \frac{1}{ \sqrt[]{1-\GeneralSuccessProb}} \sum_{\QPEbitstring \in X_\text{bad}} \GeneralSuccessProb_\QPEbitstring \ket{\varphi_\QPEbitstring} \ket{\QPEbitstring} , \quad \ket{\Phi_1}  = \frac{1}{ \sqrt[]{\GeneralSuccessProb}} \sum_{\QPEbitstring \in X_\text{good}} \GeneralSuccessProb_\QPEbitstring \ket{\varphi_\QPEbitstring} \ket{\QPEbitstring},
\end{align}
$\alpha_\QPEbitstring$ are the complex amplitudes, $\ket{\QPEbitstring}$ represents the computational basis states, and $\ket{\varphi}$ is additional workspace (the states $\ket{\QPEbitstring}$ and $\ket{\varphi}$ shall be used for the energy and state register, respectively, in the next section).  The states are orthonormal, $\braket{\Phi_i | \Phi_j} = \delta_{ij}$, and the summed probability that a measurement of $\ket{\Phi}$ yields one of the good state described by the set $X_\text{good}$ is

\begin{align}
\GeneralSuccessProb \equiv |\braket{\Phi_1 | \Phi}|^2 = \sum_{\QPEbitstring \in X_\text{good}} |\GeneralSuccessProb_\QPEbitstring|^2. \label{eq.general_ probability}
\end{align}
If $\GeneralSuccessProb \ll 1$, then the probability of observing a good state is almost zero. That is, we shall expect to repeat state preparation $O(1/\GeneralSuccessProb)$  times with $\mathcal{A}$  on average before a state with $\QPEbitstring \in X_\text{good}$ is found. The amplification process, originally proposed in Grover's database searching quantum algorithm\cite{Grover1996, Grover1997} and later revised by Brassard \emph{et. al.}\cite{Brassard2000}, improves the scaling $O(1/\alpha)$ by amplifying the amplitudes associated with the good states by repeatedly applying the following unitary operator

\begin{align}
Q(\mathcal{A}, \GoodBitVectorgeneral) = -\mathcal{A} S_0 \mathcal{A}^{-1} S_{\GoodBitVectorgeneral}. \label{Eq.general_ampli_ampli}
\end{align}
The operator $S_{\GoodBitVectorgeneral}$ conditionally changes the sign of the amplitudes of states with $x \in X_\text{good}$,

\begin{equation}
S_{\GoodBitVectorgeneral} \ket{\Phi} = \sqrt[]{1-\GeneralSuccessProb}\ket{\Phi}_0 - \sqrt[]{\GeneralSuccessProb}\ket{\Phi_1}, \label{eq.general_mark_bits}
\end{equation}
that is, it acts as an oracle which recognizes the good states. The operator $S_0$ changes the sign of the amplitude if and only if all the qubits are in the zero state $\ket{0}$. The quantum amplitude amplification algorithm is a generalization of the Grover's algorithm, in the sense the unitary $\mathcal{A}$ is not restricted to create an equal superposition in the computational basis. More details about Grover's algorithm and the comparison with the amplification process is given in \ref{App:Grovers_search_algorithm}. Following Boyer \emph{et al.}\cite{Boyer1998}, the number of times we should apply $Q(\mathcal{A}, \GoodBitVectorgeneral) $ is given by the formula 

\begin{align}
k =\bigg \lfloor \frac{\pi}{4 \arcsin(\sqrt[]{\GeneralSuccessProb})}\bigg \rfloor, \label{Eq.general_ iterations}
\end{align}
where $\alpha$ is the initial success probability given in equation (\ref{eq.general_ probability}), and it achieves a scaling of $O(1/\sqrt[]{\GeneralSuccessProb})$. Then the probability that a measurement of $Q^k(\mathcal{A}, \GoodBitVectorgeneral) \ket{\Phi}$ yields a state with $\QPEbitstring  \in X_\text{good}$ is 

\begin{align}
\text{Prob}(\QPEbitstring  \in X_\text{good})\geq \max(1-\GeneralSuccessProb,\GeneralSuccessProb), \label{Eq.general_ prob_bound}
\end{align}
where $ \max(1-\GeneralSuccessProb,\GeneralSuccessProb)$ is the lower bound of the probability. For example, consider the two scenarios: $\GeneralSuccessProb \ll 1$ or $\GeneralSuccessProb > \frac{1}{2}$. For  $\GeneralSuccessProb \ll 1$, we would have after \emph{k} iterations

\begin{align}
Q^k(\mathcal{A}, \GoodBitVectorgeneral)\ket{\Phi} \approx \sum_{x \in X_\text{good}}  \beta_x \ket{\varphi_x} \ket{x} \label{Eq. QQAgoodstates}
\end{align}
with some amplified coefficients $ \beta_x $. If $\GeneralSuccessProb> \frac{1}{2}$, we have $k = 0$ and the probability is simply the initial success probability, $\GeneralSuccessProb$. That is for $\GeneralSuccessProb > \frac{1}{2}$ it cannot be amplified further. We refer the reader to \cite{Brassard2000} which holds the proof of equations (\ref{Eq.general_ iterations}) and (\ref{Eq.general_ prob_bound}).

\subsection{Quantum amplitude estimation and Qsearch}
\label{sec:AeQs}

\subsubsection{Quantum amplitude estimation.}
\label{subsubsec: Amplitude estimation}

Overshooting the optimal \emph{k}, given in equation (\ref{Eq.general_ iterations}), may decrease the success probability. Thus knowing the probability that a measurement yields a good state is important. The quantum amplitude estimation (QAE) algorithm is an application of QPE to estimate the initial probability of success, equation (\ref{eq.general_ probability}), by estimating eigenvalues of the unitary $Q(\mathcal{A}, \GoodBitVectorgeneral)$\cite{Brassard2000}. Let $\GeneralSuccessProb = \sin^2(\theta_\GeneralSuccessProb)$ for $0 \leq \theta_\GeneralSuccessProb \leq \pi / 2$, then the action of $Q(\mathcal{A}, \GoodBitVectorgeneral)$ in matrix notation, spanned by the basis $\{\ket{\Phi_0}, \ket{\Phi_1}\}$, is given by

\begin{align}
 \underbrace{\begin{bmatrix}
       \cos(2\theta_\GeneralSuccessProb) &  -\sin(2\theta_\GeneralSuccessProb)       \\[0.3em]
        \sin(2\theta_\GeneralSuccessProb) &  \cos(2\theta_\GeneralSuccessProb)        
     \end{bmatrix}}_{Q(\mathcal{A}, \GoodBitVectorgeneral)} \underbrace{\begin{bmatrix}
       \cos(\theta_\GeneralSuccessProb)     \\[0.3em]
        \sin(\theta_\GeneralSuccessProb)      
     \end{bmatrix}}_{ \ket{\Phi}} =  \begin{bmatrix}
       \cos(3\theta_\GeneralSuccessProb)     \\[0.3em]
        \sin(3\theta_\GeneralSuccessProb)      
     \end{bmatrix}, \label{Eq. MatrixNotation}
\end{align}
where $\cos(\theta_\GeneralSuccessProb)$ and $\sin(\theta_\GeneralSuccessProb)$ are the initial amplitudes for failure and success, respectively. The amplitude amplification process boosts the angle $\theta_\GeneralSuccessProb$  to three times its original value, thereby changing the amplitude of $\ket{\Phi_1}$ to $\sin(3\theta_\GeneralSuccessProb)$ and $\ket{\Phi_0}$ to  $\cos(3\theta_\GeneralSuccessProb)$, which for  $\theta_\GeneralSuccessProb< \pi /4$ (i.e. $\GeneralSuccessProb < 1/2$) corresponds to an increase in the amplitude of the good state $\ket{\Phi_1}$. The eigenvalues of $Q(\mathcal{A}, \GoodBitVectorgeneral)$ are $e^{\pm 2 i \theta_\GeneralSuccessProb}$, and can be estimated using the QPE method. The result of QAE is a \emph{t}-bit approximation to $\theta_\GeneralSuccessProb$, and the error in our estimate $\tilde{\theta}_\GeneralSuccessProb$ for $\theta_\GeneralSuccessProb$ translates to an error in our estimate $\tilde{\GeneralSuccessProb} = \sin^2(\tilde{\theta}_\GeneralSuccessProb)$, given by

\begin{align}
| \GeneralSuccessProb  - \tilde{\GeneralSuccessProb}| \leq  2\pi \hspace{0.07cm} \QPEerrorTolerance  \hspace{0.07cm} \frac{\sqrt[]{\GeneralSuccessProb(1-\GeneralSuccessProb)}}{2^t}  + \QPEerrorTolerance^2  \hspace{0.07cm} \frac{\pi^2}{2^{2t}} \label{Eq. QAEerror}
\end{align}
with probability at least $8 /\pi^2$ when $\QPEerrorTolerance = 1$ and with probability greater than $1-\frac{1}{2(\QPEerrorTolerance-1)}$ for $\QPEerrorTolerance \in \mathbb{Z}^+/\{1\}$. Thus the QAE achieves a scaling of $O(1/2^t)$ of the estimation error. We refer the reader to \cite{Brassard2000} which develops the proof of equation (\ref{Eq. QAEerror}).  In the recent months, a family of novel related algorithms with similar goals have been proposed in  \cite{grinko2019, suzuki2020,  wie2019, aaronson2020}. For example, the iterative quantum amplitude estimation algorithm\cite{grinko2019} does not rely on QPE, and requires significantly fewer control gates and qubits to estimate the amplitude. An alternative approach is the fixed-point quantum search that avoids overshooting the optimal \emph{k}-value, and still achieves a quadratic advantage over classical unordered search\cite{yoder2014}. In the next section, we summarize the \emph{Qsearch} algorithm, based on Boyer \emph{et al.}\cite{Boyer1998}, which foregoes amplitude estimation entirely, finding a solution in expected runtime in $O(\frac{1}{\sqrt[]{\GeneralSuccessProb}})$, and does not need additional registers or control operations.

\subsubsection{Qsearch.}
\label{sec.pre.Qsearch}

The Qsearch algorithm, original proposed by Boyer \emph{et al.}\cite{Boyer1998} for the Grover search algorithm and later revised by Brassard \emph{et al.}\cite{Brassard2000} to general unitaries, finds a solution \emph{without} estimation the amplitude. The Qsearch protocol  randomly picks an integer \emph{i} and applies $Q^i(\mathcal{A}, \GoodBitVectorgeneral)$, and increases the search space exponentially for each loop. The method works as follows: Let $0 \leq i< l$, then the size of the search space is defined as \emph{l}. The probability that a measurement yields a good state after \emph{i} iterations of  $Q(\mathcal{A}, \GoodBitVectorgeneral)$ is $\sin^{2}((2i+1)\theta_\GeneralSuccessProb)$, as shown in equation (\ref{Eq. MatrixNotation}). Picking an integer \emph{i} uniformly at random such that $0 \leq i< l$, the average success probability is 

\begin{align}
\sum^{l-1}_{i = 0} \frac{1}{l} \sin^{2}((2i+1)\theta_\GeneralSuccessProb) = \frac{1}{2} - \frac{\sin(4l \theta_\GeneralSuccessProb)}{4l \sin(\theta_\GeneralSuccessProb)}. \label{eq.QsearchGeneral1}
\end{align}
If $l \geq 1/\sin(2 \theta_\GeneralSuccessProb)$ then $ \frac{\sin(4l \theta_\GeneralSuccessProb)}{4l \sin(\theta_\GeneralSuccessProb)} \leq 1/4$, and we define $l_0 :=1/\sin(2 \theta_\GeneralSuccessProb) $ as the \emph{critical stage},  meaning if $l > l_0$ the average failure probability is $\leq 3/4$. The expected total number of iterations needed to reach the critical stage, if it is reached, is at most

\begin{align}
\frac{1}{2} \sum^{\lceil \log_g l_0 \rceil}_{s = 1} g^{s-1} < 4 l_0, \label{eq.QsearchGeneral2}
\end{align}
where \emph{g} is the growth factor and set to $8/7$, and since  $0 \leq i< l$  is chosen uniformly at random we obtain the  1/2-factor. We set $g = 8/7$ to obtain a small constant '4'. Any value of \emph{g} strictly between 1 and 4/3 is allowed, however, for $g \rightarrow 1$ or $g \rightarrow 4/3$, the involved constants are very large, and render this algorithm impractical. If the critical stage is reached, the average failure probability for each loop is $\leq 3/4$. The expected total number of iterations before success after the critical stage is reached is

\begin{align}
\frac{1}{2} \sum^{\infty}_{u = 0} \bigg( \frac{3}{4}\bigg)^u g^{\lceil \log_g l_0 \rceil + u} <4l_0. \label{eq.QsearchGeneral3}
\end{align}
Thus the expected \textbf{total} runtime is less that $8 l_0 $ where $l_0 = 1/\sin(2 \theta_\GeneralSuccessProb) $ in units of $ Q(\mathcal{A}, \GoodBitVectorgeneral)$, and provided the initial success probability is very small, $\GeneralSuccessProb \ll 1$, we find $8  l_0 \approx 4 \frac{1}{\sqrt[]{\GeneralSuccessProb}} = O(\frac{1}{\sqrt[]{\GeneralSuccessProb}})$. We refer the reader to section 6 in \cite{Boyer1998} for more details about the derivations of (\ref{eq.QsearchGeneral1})-(\ref{eq.QsearchGeneral3}).

\section{Amplitude amplification of eigenstate selection}
\label{sec: our_method}
We develop two main results: First, we propose \emph{the phase-estimation interval target energy readout}, or PHILTER, which amplifies energy eigenstates within a target energy-interval. The idea of amplifying eigenstates based on QPE was first pointed out by Ammar Daskin in the context of principal component analysis\cite{Daskin2016}. Our work includes the effect of the QPE amplitudes on the amplification process. We propose an iterative version of the PHILTER algorithm, which reduces the number of qubits in the amplification process with the cost of being iterative. Second, we propose the \emph{QPHILTER} protocol, which obtains a better scaling compared to PHILTER but may be less practical. We summarize the computational complexity of each proposed algorithm in table \ref{table. complexity}.

\setlength{\tabcolsep}{12pt}
\begin{table}[H]
\centering
\caption{This table gives a summary of the different scaling of the algorithms proposed in this paper. The computational complexity is in units of $\text{QPE}(\mathcal{H})O$ for measuring an energy with $E \in \good$ . The initial success probability, \emph{b}, is defined in equation (\ref{eq. true prob}). First, QPE (second column) is the expected number of $\text{QPE}(\mathcal{H})O$'s without the amplification protocol. The PHILTER and iterative PHILTER protocol both depend on the amplitude estimation prior to amplitude amplification, which scales as $O(2^t)$. This turn out to be the dominated step in the scaling. What is the ideal value of \emph{t}? If $ \TrueProbOverlap > 0.25$ then $t_{\text{ideal}} = 2$, if $\TrueProbOverlap > 0.125$ then $t_{\text{ideal}} = 3$ etc. In general if $\TrueProbOverlap > 2^{-t}$ then $t_{\text{ideal}} = \lceil \log_2(1/\TrueProbOverlap) \rceil$. We cannot determine $t_{\text{ideal}}$ since it depends on the unknown initial success probability. In practice the possible choices of \emph{t} will be limited by the available quantum hardware. The pragmatic approach would then be to use the largest \emph{t} value possible. Physical insight about the individual problem at hand combined with heuristics might help in choosing the \emph{t} value in future applications.}
\begin{tabular}{l*{6}{c}r}
\br
Algorithm            &Scaling& &   \\
\hline
QPE& $O(1/\TrueProbOverlap)$    & section \ref{sec. QPE}   \\[0.2cm]
 PHILTER  &  $O(2^{\lceil \log_2(1/\TrueProbOverlap) \rceil})$ & section \ref{sec: PHILTER}, figure \ref{fig. FILTERalg}, algorithm \ref{AlgPHILTER} \\[0.2cm]
  Iterative PHILTER  &$O(2^{\lceil \log_2(1/\TrueProbOverlap) \rceil})$  & section \ref{sec:iaa}, figure \ref{fig. Iterative_amplitude_amplification}  \\[0.2cm]
  \textsf{Q}PHILTER  & $O(1/\sqrt[]{\TrueProbOverlap})$ &section  \ref{sec:Qsearch_PHILTER}, algorithm \ref{AlgQsearch} \\[0.2cm]
\br
\end{tabular}
\label{table. complexity}
\end{table}

\subsection{PHILTER: The phase-estimation interval target energy readout}
\label{sec: PHILTER}
\begin{figure}[H] 
\centering  
\includegraphics[width=0.6\textwidth]{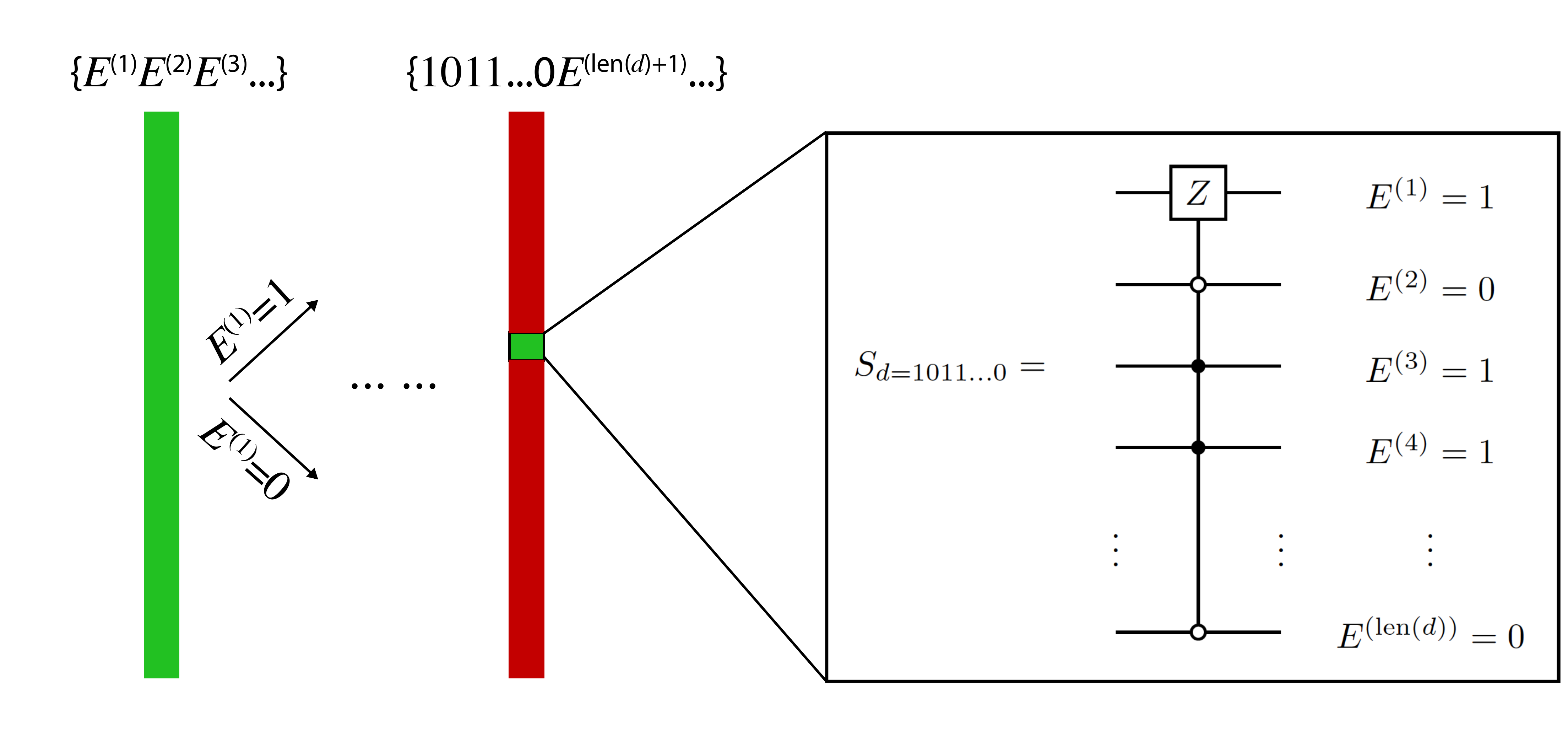} 
\caption{ Schematic representation of marking the target energy-interval. Shown here is how to construct the gate $S_{\GoodBitVector}$ which recognizes the good energy eigenstates, $E \in \good$. The gate applies a phase on the bit-string $\GoodBitVector$, in this case $\GoodBitVector = 1011\hdots 0$. The length of the string $\GoodBitVector$, denoted as $\LenGoodBitVector$, satisfies $\LenGoodBitVector \leq m$ and controls the size of the target energy-interval, i.e. the amplification process amplifies the $\LenGoodBitVector$ most significant bits in the binary representation of the energy.  }
\label{fig. oracle}
\end{figure}

Consider equation (\ref{Eq. algorithm}) split in the same way as equation (\ref{Eq.general_QAAstate}), where $\mathcal{A} = \text{QPE}(\mathcal{H})\emph{O}$ and the good states are those that have energies within the target energy-interval, $E \in  \good  $. Thus we write equation (\ref{Eq. algorithm}) as the superposition of all the good states, $\ket{\Psi_1}: E \in \good$, and all the bad states $\ket{\Psi_0}: E \notin \good$ of $\ket{\Psi}$, where

\begin{align}
\ket{\Psi_1} = \frac{1}{\sqrt[]{a}}\sum_{ E_j \in \good } \AnsatzProbOverlap_j \ket{E_j} \bigg(\sum^{2^m}_{\QPEbitstring_i \in \{0,1\}^m} \epsilon_{\QPEbitstring_i}^{(j)} \ket{\QPEbitstring_i} \bigg) \approx \frac{1}{\sqrt[]{a}}\sum_{ E_j \in \good } \AnsatzProbOverlap_j \ket{E_j} \ket{E_j^{(1)}E_j^{(2)}\hdots E_j^{(m)}}, \label{eq. good states}
\end{align}
$E_j^{(1)}E_j^{(2)}\hdots E_j^{(m)}$ is the best \emph{m}-bit approximation to the energy eigenvalue, the states are orthonormal, $\braket{\Psi_i | \Psi_j} = \delta_{ij}$, and

\begin{align}
\AnsatzProbOverlap \equiv |\braket{\Psi_1 | \Psi}|^2 = \sum_{E_j \in \good} |\AnsatzProbOverlap_j|^2 \label{eq. probability}
\end{align}
denotes the probability that a measurement of $\ket{\Psi}$ yields a good state. Equation (\ref{eq. good states}) assumes that QPE produces only the best \emph{m}-bit approximation to the energy eigenvalues in the energy register --- this is only true if the energy eigenvalues can be  exaclty represented with \emph{m}-bits. We will address the effect of the QPE amplitudes later in this section, however, if  \emph{m} is large the approximation is valid due to the suppression of the erroneous QPE amplitudes. The value of equation (\ref{eq. probability}) depends on the prepared ansatz. For example, if $\AnsatzProbOverlap \ll 1$, meaning the overlap between the prepared ansatz and the good energy eigenstates is very small, then the probability of observing a good state is almost zero using the QPE method. That is, we shall expect to repeat state preparation  with $\text{QPE}(\mathcal{H})O$  $O(1/\AnsatzProbOverlap )$  times on average before a state with $E \in \good$ is found. Here, we want to improve the scaling  $O(1/\AnsatzProbOverlap)$ by amplifying the amplitudes associated with the good states. The amplitude amplification process, equation (\ref{Eq.general_ampli_ampli}), allows the usage of less targeted initial trial states. The operator $S_{\GoodBitVector}$ is chosen such that it conditionally changes the sign of the amplitudes of states with $E \in \good$,  as shown in equation (\ref{eq.general_mark_bits}), and $\GoodBitVector$ denotes the first bits chosen to mark the target energy-interval. For example, say we want energies with either $E^{(1)} = 0$ or $E^{(1)} = 1$, where $E^{(1)}$ is the most significant bit in the binary representation of the energy. Then the string of specified bits, $\GoodBitVector$, is $\GoodBitVector = 0$ for $E^{(1)} = 0$ or  $\GoodBitVector = 1$ for $E^{(1)} = 1$. This put you firmly in the $E \geq 0.5$ or $E < 0.5$ interval, as depicted in figure \ref{fig. SchematicRep}B. The operator $S_{\GoodBitVector}$ is then very simple - it is just a phase applied to the first qubit of the energy register $\ket{E^{(1)}E^{(2)}\hdots E^{(m)}}$. Thus if we want energies where $E^{(1)} = 0$ then we apply the Pauli string $XZX$ on the first qubit on the energy register, or we apply \emph{Z} if $E^{(1)} = 1$ is wanted. As a result we can diminish the amplitudes related to the "wrong" value of $E^{(1)}$ using the method of equation (\ref{Eq.general_ampli_ampli}). Say we want energies with either $E^{(1)}E^{(2)} = 00 \lor 01 \lor 10 \lor 11$, i.e. we have $\GoodBitVector =  00 \lor 01 \lor 10 \lor 11$. This put you firmly in one of the four smaller energy intervals, as depicted in figure \ref{fig. SchematicRep}B. In this case, the operator $S_{\GoodBitVector}$ is then a 2-qubit gate applying a phase on the good states. In general, we may construct the operator $S_{\GoodBitVector}$ as shown in figure \ref{fig. oracle},
 
\hfill \break
thereby recognizing the good states $E \in \good$ by applying a phase on these states. The length of the amplified bitstring $\GoodBitVector$, denoted as $\LenGoodBitVector$, satisfies $\LenGoodBitVector \leq m$ and controls the size of the fixed target energy-interval, i.e. the amplification process amplifies the $\LenGoodBitVector$ most significant bits in the binary representation of the energy. The operator $S_{\GoodBitVector}$, as depicted in figure \ref{fig. oracle}, can be implemented with use of $(\LenGoodBitVector-1)$ working qubits and $2(\LenGoodBitVector -1)$ Toffoli gates, where each Toffoli gate can be implemented using Hadamard, phase, controlled-\texttt{NOT} and $\pi/8$  gates\cite{Chuang2011,Yu2015, Yu2013, jones2013, Biswal2019}. Finally, the operator $S_0$ changes the sign of the amplitude if and only if all the qubits are in the zero state $\ket{0}$, and may be implemented as in figure \ref{fig. oracle}. 

Unsuccessful amplification,  i.e. amplifying unwanted energy eigenstates, may occur because

\begin{itemize}
\item[$\bullet$] Energy eigenvalues in general cannot be written exactly with \emph{m}-bits, thus there is a small portion of unwanted energy eigenstates associated with $\GoodBitVector $, as depicted in figure \ref{fig. DE}A.
\item[$\bullet$] The initial success probability can be very small and comparable to the QPE amplitudes.
\end{itemize}
In the following we will bound the number of qubits needed in the energy register, $\ket{E^{(1)}E^{(2)}\hdots E^{(m)}}$, to guarantee successful amplification. We refer the reader to \ref{app.stabilizer-qubits}  for a more detailed explanation. In equation (\ref{eq. good states}) we assumed error-free QPE in the sense that bit-strings in $\GoodBitVector $ only resulted from the good energy eigenstates.  The full effect of QPE, however, takes the general form in equation (\ref{Eq. algorithm}), and defining the set:

\begin{figure}[t]  
\centering
\includegraphics[width=1.0\textwidth]{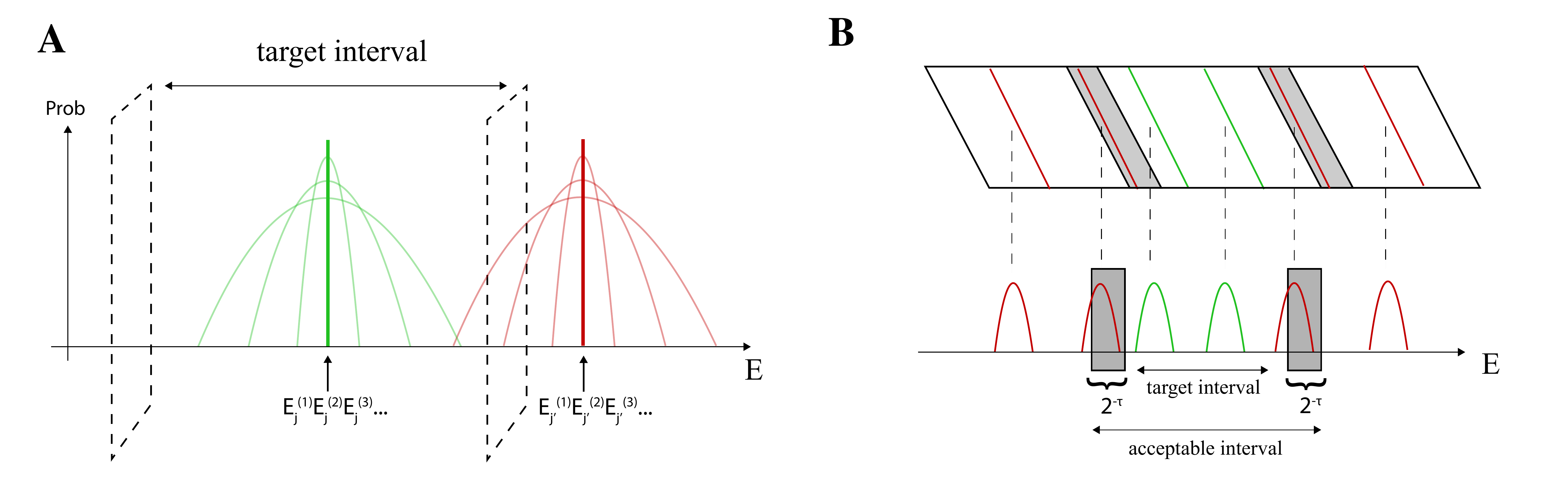} 
\caption{Schematic depiction of the spread in QPE amplitudes $\epsilon^{(j)}_{\QPEbitstring_i}$, and the effect of this on the amplification process. \textbf{A}: The probability distributions reflect the spread in QPE amplitudes and is centred around the energy eigenvalues, $E_j^{(1)}E_j^{(2)}E_j^{(3)}\hdots$.  Increasing the number of qubits in the energy register narrow down the width of the distributions.   \textbf{B}: Schematic representation of accepting energy eigenvalues outside the target interval, but within the larger interval (grey area). }
\label{fig. DE}
\end{figure}

\begin{align}
\text{Good QPE-outputs:} \quad \big\{\text{$\ket{\QPEbitstring_i} \big|$ $\QPEbitstring_i \in \GoodSet$} \big\},
\end{align}
where the ``good QPE-outputs'' start with fixed bit values given by the bitstring $\GoodBitVector$, depicted in figure \ref{fig. oracle}, and described by the set $ \GoodSet$. We again write equation (\ref{Eq. algorithm}) in the form of equation (\ref{Eq.general_QAAstate}), formally splitting it up into the summed good and bad states, where the good state $\ket{\Omega_1}$ is given by

\begin{align}
\ket{\Omega_1} = \frac{1}{\sqrt[]{\TrueProbOverlap}}\sum_{j} a_j \ket{E_j} \bigg(\sum_{\QPEbitstring_i \in \GoodSet} \epsilon_{\QPEbitstring_i}^{(j)} \ket{\QPEbitstring_i} \bigg),
\end{align}
and

\begin{align}
\TrueProbOverlap \equiv  |\braket{\Omega_1 |\Psi }|^2 = \sum_{\substack{E_j \in \good, \QPEbitstring_i \in \GoodSet}} |\AnsatzProbOverlap_j \epsilon^{(j)}_{\QPEbitstring_i}|^2 +  \sum_{\substack{E_j \notin \good,\QPEbitstring_i \in \GoodSet}} |\AnsatzProbOverlap_j \epsilon^{(j)}_{\QPEbitstring_i}|^2 \label{eq. true prob}
\end{align}
denotes the probability that a measurement of $\ket{\Psi}$  yields a good output string when including the effect of QPE amplitudes. Running the amplification algorithm, the states that we amplify are those that after QPE results in the correct output bit-strings $ \QPEbitstring \in \GoodSet$. For the amplification process to be successful, i.e. to reduce unwanted energy eigenstates, the following condition must be met

\begin{align}
\sum_{\substack{E_j \in \good, \QPEbitstring_i \in \GoodSet}} |\AnsatzProbOverlap_j \epsilon^{(j)}_{ \QPEbitstring_i}|^2  > \sum_{\substack{E_j \notin \good,  \QPEbitstring_i \in \GoodSet}} |\AnsatzProbOverlap_j \epsilon^{(j)}_{ \QPEbitstring_i}|^2, \label{eq. true_condition}
\end{align}
otherwise the portion of unwanted energy eigenstates associated with $\GoodBitVector$ is greater than the wanted ones, and the probability would be greater to obtain an unwanted energy eigenstate after the amplification process than a wanted one. For example, consider the case where an energy eigenvalue is just outside the target interval, as depicted in figure \ref{fig. DE}A. In that case, a significant portion of the unwanted state is associated with $\GoodBitVector$, due to the QPE amplitudes. Increasing the number of qubits in the energy register would decrease the portion of the unwanted state to a point where we meet the condition (\ref{eq. true_condition}). In practice, we cannot determine when the above condition is met since it depends on the spectrum of the Hamiltonian. However, if we agree to accept also energies sufficiently close to the target interval, i.e. energies within a larger interval $\good'>\good$, the required number of qubits can be bounded. Specifically, we define the acceptance interval  $\good'$ as the interval $\good$ expanded a distance $2^{-\PHILTERtolerance}$ to either side, where $1\leq \PHILTERtolerance $ with $\PHILTERtolerance \in \mathbb{R}$ is a parameter which determines our tolerance for error.  Then for successful amplification we need at least $ \LenGoodBitVector + \stabilizer$ qubits in the energy register, where

\begin{figure}[t]  
\centering
\includegraphics[width=0.8\textwidth]{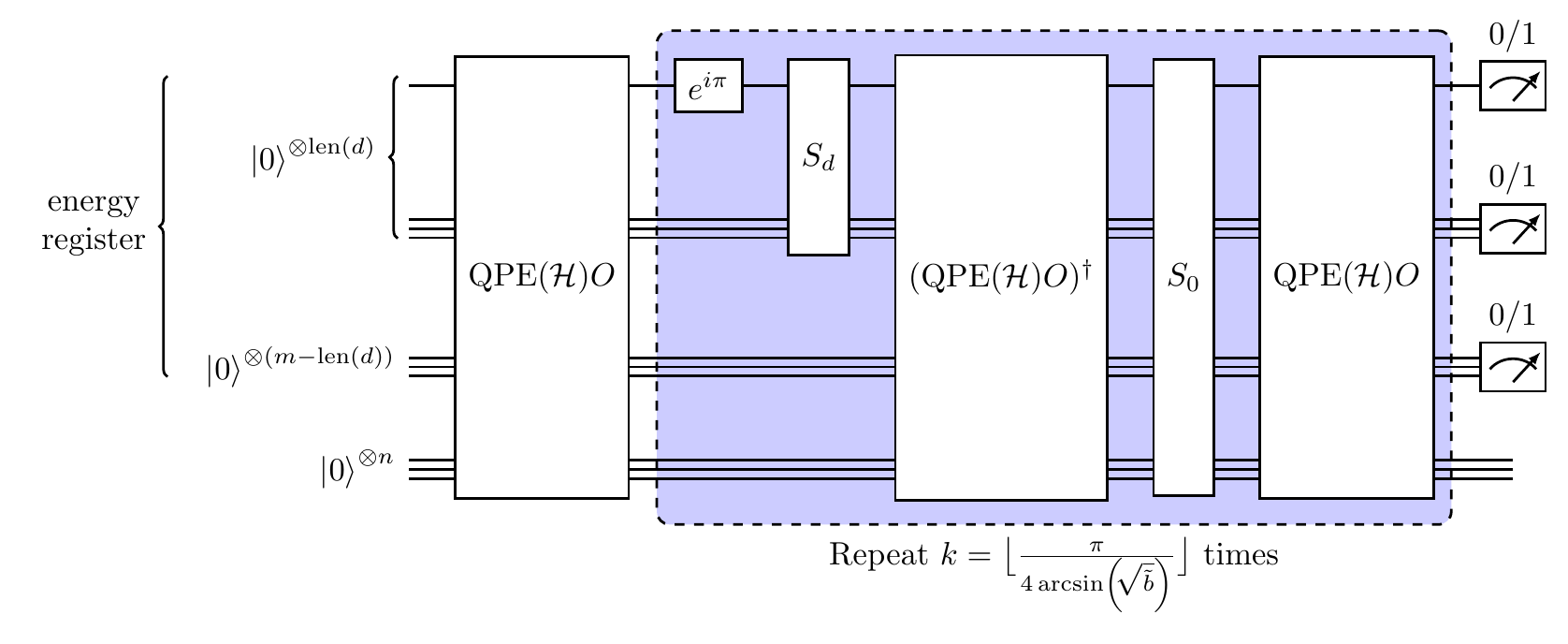} 
\caption{Circuit representation of PHILTER. A series of amplitude amplification processes, $Q(\GoodBitVector)= - \mathcal{A}S_0\mathcal{A}^{-1}S_{\GoodBitVector}$, where $\mathcal{A} = \text{QPE}(\mathcal{H}) \emph{O}$ and $S_{\GoodBitVector}$ recognize the target energy-interval, which amplify the target energy eigenstates. We repeat this process \emph{k} times to achieve a probability at least $\max(1-b,b)$ for a measurement to yield a state within the target energy-interval, $E \in  \good $, where \emph{b} is the initial success probability of $\mathcal{A}\ket{0}^{\otimes (m+n)}$. We need $m+n$ qubits, where \emph{m} determines the precision of the binary representation of the energy and \emph{n} is the number of qubits necessary to store the ansatz $O\ket{0}^{\otimes n}$. }
\label{fig. FILTERalg}
\end{figure}

\begin{align}
 \stabilizer = \bigg \lceil \log_2 \big( 2/\TrueProbOverlap\big)  + \PHILTERtolerance- \LenGoodBitVector  \bigg \rceil\label{eq.iaa-qubits}
\end{align}
 and \emph{b} is the initial success probability given in equation (\ref{eq. true prob}). That is  the amplification protocol may not filter out energy eigenvalues outside the target interval $\good$ but within the larger interval $\good'$ increased by $\pm2^{-\PHILTERtolerance}$, as depicted in figure \ref{fig. DE}B. We refer the reader to \ref{app.stabilizer-qubits} which holds the proof of equation (\ref{eq.iaa-qubits}).

The algorithm naturally divides into two pieces. First, step 1 computes an estimation of equation (\ref{eq. true prob}) using e.g. the QAE algorithm in section \ref{subsubsec: Amplitude estimation}. Next, step 2 amplifies the target energy eigenstates and output an energy eigenvalue within the target interval.  In units of the standard QPE algorithm, i.e. applying the unitary $\text{QPE}(\mathcal{H})\emph{O}$, the runtime of estimating the initial success probability using QAE scales as $O(2^{t})$ in a series and controlled by the \emph{t}-register, in itself an expensive circuit due to the many additional controlled gates. Once the amplitude has been estimated, running the algorithm requires $2\tilde{k} + 1$ further QPE steps, where $\tilde{k}$ is given in equation (\ref{Eq.general_ iterations}) given the estimate of the amplitude. The pseudocode and circuit representation for the PHILTER protocol are given in algorithm \ref{AlgPHILTER} and figure \ref{fig. FILTERalg}, respectively.

\hfill \break
\hfill \break
\fbox{
 \addtolength{\linewidth}{-2\fboxsep}%
 \addtolength{\linewidth}{-5\fboxrule}%
\begin{algorithm}[H]
\textbf{Algorithm: PHILTER} \\
\textbf{Inputs}: $\mathcal{H}$: Hamiltonian of interest, $O$: initial state preparation, $\GoodBitVector \in \{0,1\}^{\LenGoodBitVector}$: target energy-interval, $\PHILTERtolerance \in \mathbb{R} | \PHILTERtolerance \geq 1$: error tolerance and $t \in \mathbb{Z}^+ $: \emph{t}-bit precision of the initial amplitude.  \\
\textbf{Output}: An estimate of $E \in \good'$ with precision $2^{-m}$ on the energy eigenvalue.
\hfill \break
\begin{itemize}[leftmargin=*]
 \setlength\itemsep{0.1em}
 \item[1.] \textbf{Amplitude estimation}.
 \begin{itemize}
 \item[(a)] $\tilde{\TrueProbOverlap}$ $\leftarrow$ An estimate of the initial success probability $\TrueProbOverlap$ with precision $2^{-t}$ \\ (section \ref{subsubsec: Amplitude estimation}) 
    \setlength\itemsep{0.5em}
  \item[(b)] $\tilde{k}$ $\leftarrow$ $\lfloor \frac{\pi}{4 \arcsin(\sqrt[]{\tilde{b}})} \rfloor $, estimate the optimal number of times we should apply $Q(\text{QPE}(\mathcal{H})O, \GoodBitVector)$
 \end{itemize}
  \setlength\itemsep{1.5em}
 \item[2.] \textbf{Amplification protocol}. Let $m \geq \lceil \log_2 \big( 2/\tilde{b}\big)  + \PHILTERtolerance \rceil$:
   \setlength\itemsep{0.5em}
  \begin{itemize}
  \setlength\itemsep{0.5em}
 \item[(a)] Apply $Q^{\tilde{k}}(\text{QPE}(\mathcal{H})O, \GoodBitVector)$ on $\text{QPE}(\mathcal{H})O\ket{0}^{\otimes( n +m)}$, and measure the energy register (\emph{m}-qubit register).
  \item[(b)] If the output is $E \in \good'$, the problem is solved: \textbf{exit}.
  \item[(c)] Otherwise, go back to step 2 (a)
 \end{itemize}
 \end{itemize}
\hfill \break
 \caption{Pseudocode for the PHILTER protocol}
 \label{AlgPHILTER}
\end{algorithm}}

\subsection{Iterative PHILTER}
\label{sec:iaa}

\begin{figure}[t]  
\centering
\includegraphics[width=0.9\textwidth]{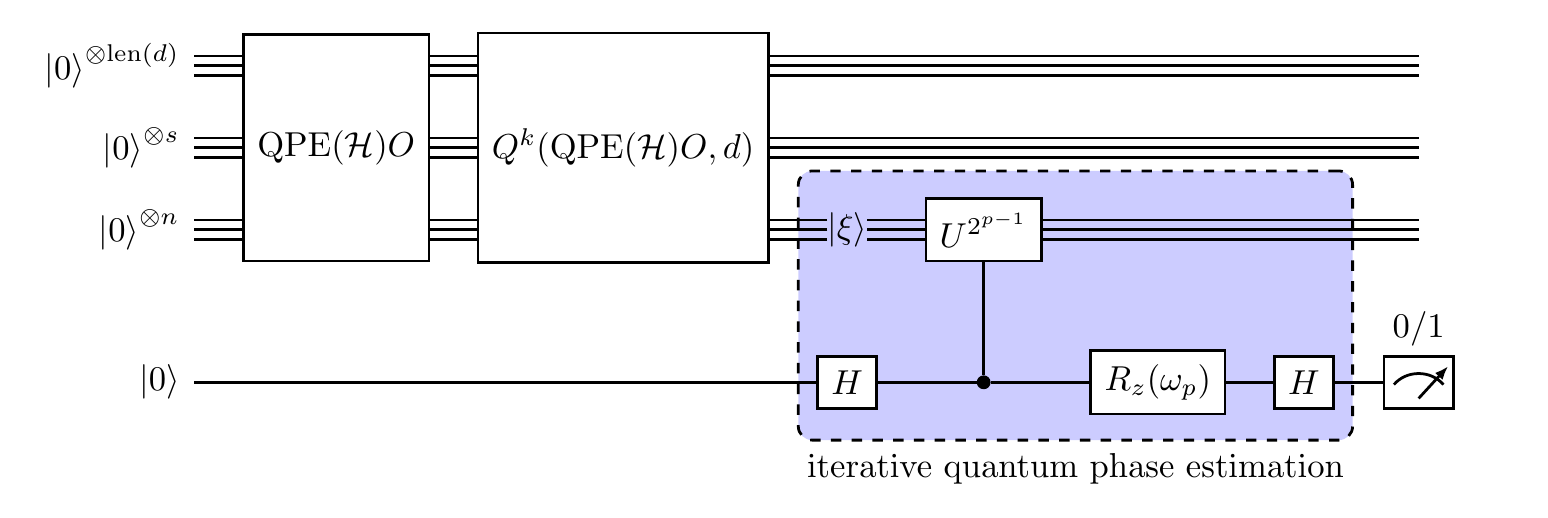} 
\caption{ Circuit representation of the iterative PHILTER. The first register consists of the $\LenGoodBitVector$ qubits we amplify, the second $\stabilizer$ qubits register suppresses the erroneous QPE amplitudes for successful amplification, the third \emph{n} qubits register stores the ansatz and the fourth register (a single ancillary qubit) encodes the information about the missing bits in the energy eigenvalues. We use the amplified state, $\ket{\xi}$, to extract an energy eigenvalue within the target interval. The angle $\omega_p$ depends on all previously measured bits as $\omega_p = -2\pi (0.0E^{(p+1)}E^{(p+2)} \hdots E^{(m)})$ and $\omega_m = 0$. }
\label{fig. Iterative_amplitude_amplification}
\end{figure}

We introduce now the iterative version of the PHILTER algorithm, based on iterative quantum phase estimation (IQPE). As detailed by Lanyon \emph{et al.}\cite{Lanyon2010} and Dob\v s\' i\v cek \emph{et al.}\cite{Miroslav2007}, the number of ancilla qubits can be greatly reduced compared to QPE while maintaining the same precision and probability distribution. We refer the reader to \cite{Miroslav2007} for a detailed analysis of IQPE. The PHILTER protocol above, as depicted in algorithm \ref{AlgPHILTER}, requires \emph{m} qubits in the energy register for a precision of $ 2^{-m}$ on the target energy eigenvalues. Reducing the number of qubits from \emph{m} to $\LenGoodBitVector+ \stabilizer$ when $\LenGoodBitVector+ \stabilizer \leq m$, where $\LenGoodBitVector$ controls the target energy-interval and $\stabilizer$ is given in equation (\ref{eq.iaa-qubits}), results in a lower energy precision, but we can still amplify the target energy-eigenstates. In order to maintain the same precision \emph{m}, we then incorporate the IQPE algorithm on a second register. The iterative PHILTER algorithm has four registers, as depicted in figure  \ref{fig. Iterative_amplitude_amplification}; the first register consists of the $\LenGoodBitVector$ qubits we amplify, the second $\stabilizer$ qubits register suppresses the erroneous QPE amplitudes for successful amplification, the third \emph{n} qubits register stores the ansatz and the fourth register (a single ancillary qubit) encodes the information about the missing bits in the energy eigenvalues. Immediately before the IQPE, the system state is

\begin{align}
\approx\bigg(\sum_{\substack{ E_j \in \good', \QPEbitstring_i \in \GoodSet}}  c_{\QPEbitstring_i}^{(j)} \ket{E_j} \ket{\QPEbitstring_i}  + \sum_{\substack{ E_j \notin \good', \QPEbitstring_i \in \GoodSet}}  c_{\QPEbitstring_i}^{'(j)} \ket{E_j} \ket{\QPEbitstring_i} \bigg) \otimes  \ket{0} 
\end{align}
with some amplified coefficients, $  c_{q_i}^{(j)},  c_{q_i}^{'(j)}$.  The IQPE first iteration extracts the least significant bit, $E^{(m)}$, and we repeat each iteration to obtain the correct bit. A single iteration would not be enough due to the amplified state being a linear combination of energy eigenstates. Furthermore, if the condition (\ref{eq. true_condition}) is not met, then we obtain with almost certainty an unwanted energy thus unsuccessful amplification. To ensure this not the case, we use $\LenGoodBitVector + \stabilizer$ qubits in the energy register.

\subsection{ \textsf{Q}PHILTER}
\label{sec:Qsearch_PHILTER}
Our version of the Qsearch algorithm\cite{Boyer1998} is to incorporate the QPE to search after energy eigenvalues without knowing the initial success probability. The complete algorithm is given as following:

\hfill \break
\hfill \break
\fbox{
\begin{algorithm}[H]
\textbf{Algorithm: \textsf{Q}PHILTER} \\
\textbf{Inputs}: $\mathcal{H}$: Hamiltonian of interest, $O$: initial state preparation, $\GoodBitVector \in \{0,1\}^{\LenGoodBitVector}$: target energy-interval, $\PHILTERtolerance \in \mathbb{R} | \PHILTERtolerance \geq 1$: error tolerance  and $m \in \mathbb{Z}^+: $ energy precision.  \\
\textbf{Output}: An estimate of $E \in \good$ with precision $2^{-m}$ on the energy eigenvalue.
\hfill \break
\begin{itemize}[leftmargin=*]
 \setlength\itemsep{0.1em}
 \item[1.]  Initialization;  $l = 1$ and set the growth factor $g = 8/7$
 \item[2.] Choose an integer \emph{k} uniformly at random such that $0 \leq k < l$
 \item[3.] Apply $Q^{k}(\text{QPE}(\mathcal{H})O, \GoodBitVector)$ on $\text{QPE}(\mathcal{H})O\ket{0}^{\otimes( n +m)}$, and measure the energy \\register (\emph{m}-qubit register)
 \item[4.] If the output $E$ is good, that is, if $E \in \good$, the problem is solved: \textbf{exit}
 \item[5.] Otherwise, set \emph{l} to $g\cdot l$ and go back to step 2
 \end{itemize}
 \hfill \break
 \caption{Pseudocode for the \textsf{Q}PHILTER protocol}
 \label{AlgQsearch}
\end{algorithm}}
 \hfill \break
 \hfill \break
A drawback of this method is that if the condition (\ref{eq. true_condition}) is not met, the algorithm may run forever.

\section{Numerical demonstrations}
\label{sec. numerical}

In this section, we proceed to test the ideas and methods introduced earlier with numerical simulations. We consider the molecular non-relativistic electronic Hamiltonian within the Born-Oppenheimer approximation

\begin{align}
\mathcal{H} = \sum_{PQ} h_{PQ}  a_P^\dagger a_Q + \frac{1}{2} \sum_{PQRS} h_{PQRS} a_P^\dagger  a_R^\dagger a_S  a_Q + h_{\text{nuc}} \label{Eq.SQH}
\end{align}
where $h_{PQ}$ and $ h_{PQRS}$ are one- and two-electron integrals in Dirac notation --- the one-electron integrals involving the electronic kinetic energy and the electron-nuclear attraction, and the two-electron integrals involving the electron-electron interaction. The scalar term, $h_{\text{nuc}}$, represents the nuclear-repulsion energy. 
\hfill \break
\begin{figure}[t]  
\centering \includegraphics[width=1.0\textwidth]{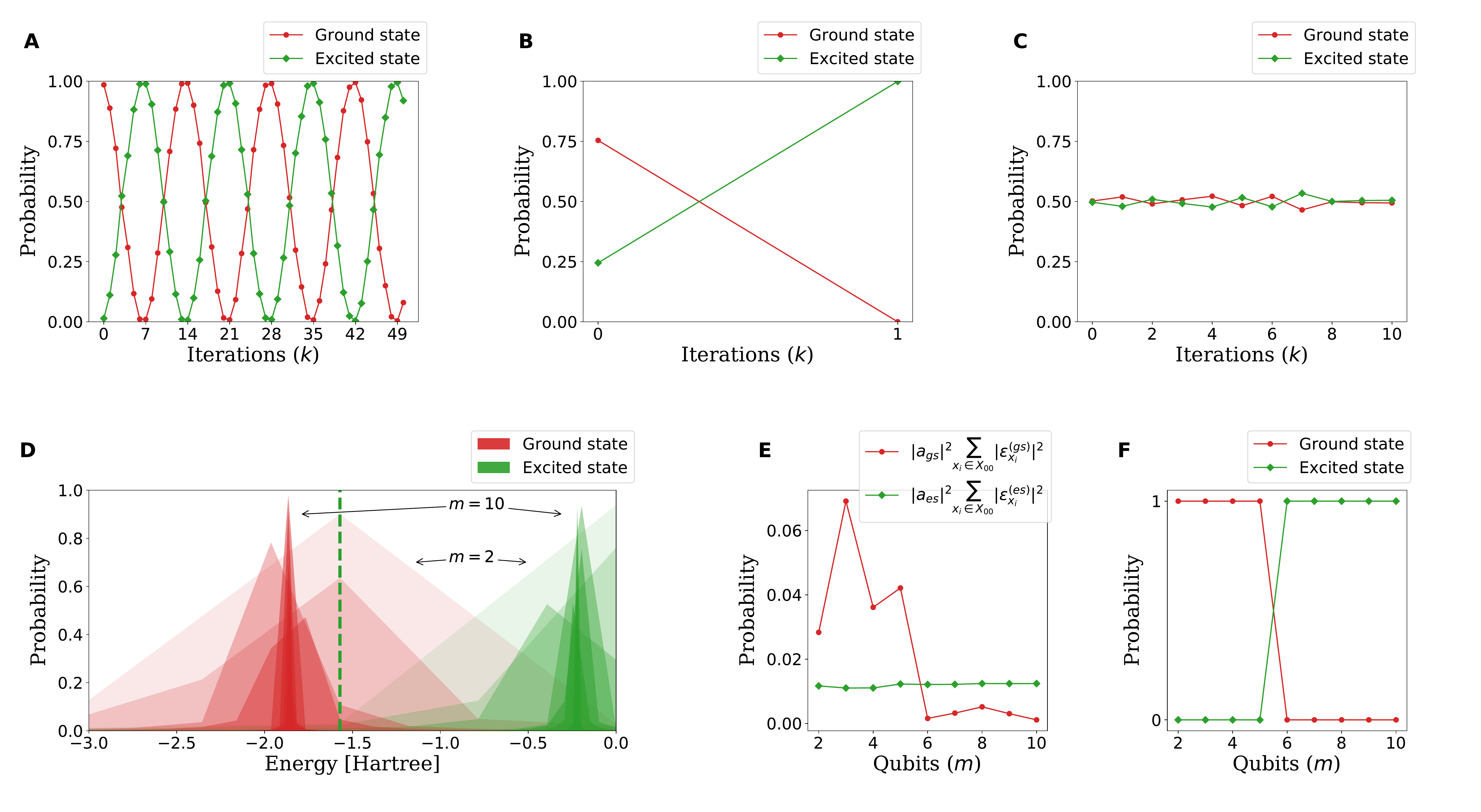} 
 \caption{PHILTER protocol for molecular Hydrogen in two spatial orbitals (STO-3G). We use the Cirq simulator\cite{ho2018} to simulate noiseless quantum circuits. The time-evolution operator $\exp(i\mathcal{H}^{(1)}t)$, obtained at the equilibrium bond length 0.7348$\text{\r{A}}$, with $\mathcal{H}^{(1)}\in \text{dim(2)}$ is encoded exactly. The energy register (\emph{m}-qubit register) of the quantum phase estimation (QPE) algorithm contains 20 qubits. \textbf{A}: The probability that a measurement yields the ground or excited state energy as a function of iterations (\emph{k}), i.e. $|\bra{E^{(1)}\hdots E^{(20)}}Q^k(  \GoodBitVector =00)\ket{\Psi}|^2$ where $\ket{\Psi} = \text{QPE}(\mathcal{H}^{(1)})R_y(\theta = 0)\ket{0}^{\otimes 20 + 1}$ and $\ket{E^{(1)}\hdots E^{(20)}}$ represents the energy register. \textbf{B}: A special case of the amplitude amplification process using the ansatz $ |\bra{E_{es}}R_y(\theta = 0.824)\ket{\text{HF}}|^2 = 1/4 $ and \textbf{C}: $ |\bra{E_{es}}R_y(\theta = 1.347)\ket{\text{HF}}|^2 = 1/2 $. \textbf{D}: Spread in QPE amplitudes $(\epsilon^{(j)}_{\QPEbitstring_i})$ for different values of $m =2, \hdots, 10$ using the $\ket{\text{HF}}$ ansatz. The target interval ($\GoodBitVector = 00$ $\rightarrow$ $\good = (-1.57,0]$) is highlighted with the dashed vertical line. \textbf{E}:  The QPE probability of the ground and excited state to output computational basis-states starting with the correct bit-string, $\GoodBitVector = 00$, described by the set $X_{00}$, as a function of number of qubits in the energy register when using the ansatz $\ket{\text{HF}}$. \textbf{F}: The probability that a measurement yields the ground or excited state energy as a function of qubits in the energy register, when running the iterative  PHILTER protocol, given the optimal number of iterations $k = 7$ and ansatz $\ket{\text{HF}}$.}
\label{fig. H2_PHILTER}
\end{figure}
\hfill \break
\emph{Molecular Hydrogen STO-3G basis.-} Consider molecular Hydrogen ($\text{H}_2$) in a minimal atomic basis (here we use STO-3G) resulting in two spatial molecular orbitals \emph{G} and \emph{U}, where \emph{G} denotes even (\emph{gerade}) and \emph{U} odd (\emph{ungerade}) inversion symmetry of the orbitals. We can write a Slater determinant in the occupation number basis as $\ket{f_{G\uparrow}f_{G\downarrow}f_{U\uparrow}f_{U\downarrow}} $, where $f_i = 1$ if spin-orbital \emph{i} is occupied, and $f_i = 0$ if spin-orbital \emph{i} is unoccupied, and the arrows denote the spin-state. Due to the high symmetries in the system we can restrict ourselves to the subspace of 2-electron singlet states in the \emph{G} representation, which is spanned by the mean-field (Hartree-Fock) configuration $\ket{\text{HF}} = \ket{1100} \equiv \ket{0} $ and the excited configuration $\ket{0011} \equiv \ket{1}$,

\begin{equation}
\mathcal{H}^{(1)}:\begin{cases}
     \ket{E_{gs}} =  -0.9938\ket{0} + 0.1115 \ket{1} \\
   \ket{ E_{es}} =  0.1115\ket{0} +  0.9938\ket{1},  
  \end{cases} \label{eq. H1}
\end{equation}
where  $\mathcal{H}^{(1)}$ is the Hamiltonian of the subspace, and $\ket{E_{gs}}$ (ground state) and $\ket{E_{es}}$ (excited state) are the energy eigenstates of $\mathcal{H}^{(1)}$ given at 0.7348$\text{\r{A}}$ proton-proton distance (equilibrium bond length) with eigenvalues $ -1.8574E_h$ and $ -0.22441E_h$ in Hartree units, respectively. Here we omit the nuclear-repulsion energy. At equilibrium bond length the ground state has a significant overlap with the HF state, $\braket{E_{gs} |\text{HF}} = 0.9938$, thus if we were to prepare the HF state for the QPE method, then we would expect to repeat QPE $O(1/(0.1115)^2) = O(80) $ times on average before the excited state energy is found. If finding the excited state is the goal, we could in this case simply prepare the state $\ket{1}$, which has a significant overlap with the excited state. Also for the system sizes considered, the energies can be found effectively through classical diagonalization. Thus, the purpose of this example is a proof-of-concept to test the ideas and methods introduced earlier. It is a useful example because we can easily generate any ansatz of interest using the compressed representation. The two possible energies of $\mathcal{H}^{(1)}$ to a precision of $2^{-20}E_h$ are given by

\begin{align}
 E_{gs} &=\underbrace{01001011101011100010}_{\text{$ E_{gs}^{(1)} E_{gs}^{(2)} \hdots E_{gs}^{(20)}$ (\text{QPE output})}} \quad \text{and} \quad E_{es} =\underbrace{00001001001001101111}_{\text{$ E_{es}^{(1)} E_{es}^{(2)} \hdots E_{es}^{(20)}$  (\text{QPE output)}}}, \label{Eq.esBitString}
\end{align}
where the bit-strings $E_{gs}^{(1)} E_{gs}^{(2)} \hdots E_{gs}^{(20)}$ and $E_{es}^{(1)} E_{es}^{(2)} \hdots E_{es}^{(20)}$ are the QPE output (with almost unity probability) when setting $m = 20$, and the rescaling factor $-2\pi E_h / 2^{20} $ converts the bit strings to energies. The bits are numbered from left to right starting from the most significant bit. The most significant bit for the two energies is identical, $E^{(1)}_{gs} = E^{(1)}_{es} = 0$, but they differ for the second most significant bit. Thus amplifying the bit string $\GoodBitVector= 00$ would amplify the amplitude associated with the excited state and reduce the probability of the ground state. In the compressed representation, we can generate any ansatz of interest using a single-qubit gate $R_y(\theta)$, for example $R_y(\theta = 0)\ket{0} = \ket{0} = \ket{\text{HF}}$, and the time-evolution operator can be decomposed into a global phase and a series of rotations of the one-qubit Hilbert space,

\begin{align}
U = \exp(-i \mathcal{H}^{(1)} t) = e^{-i\alpha} R_y(\beta) R_z(\gamma)R_y(-\beta), \label{Eq.timeevolH2}
\end{align}
where the angles are given by $\alpha = t(-1.8574 -0.22441 )/2$, $\beta = -2\arccos(0.9938)$ and  $\gamma = t(-1.8574 +0.22441 )$  at 0.7348$\text{\r{A}}$ proton-proton distance. The exact ground and excited states were computed using exact diagonalization to obtain the angles. Using the ansatz $\ket{\text{HF}}$, the probability that a measurement yields the excited state energy is given by $|\braket{E_{es} | \text{HF}}|^2 = 0.0124$. Our goal is to increase the probability by amplifying the bit string $\GoodBitVector = 00$ using the method in (\ref{Eq.general_ampli_ampli}). First, in figure \ref{fig. H2_PHILTER}, we show numerical data for the PHILTER protocol (algorithm \ref{AlgPHILTER}). In figure \ref{fig. H2_PHILTER}A, for $k = 0$, the probability is the initial success probability, $\AnsatzProbOverlap =0.0124 $, where $\AnsatzProbOverlap$ is defined in equation (\ref{eq. probability}). We can assume $\AnsatzProbOverlap  \approx \TrueProbOverlap$, where $ \TrueProbOverlap$ is defined in equation (\ref{eq. true prob}), since we are using 20 qubits in the energy register meaning a negligible portion of the unwanted  ground state is associated with the target interval $(\GoodBitVector = 00)$. The optimal \emph{k}-value can be determined by the formula

\begin{align}
k  =\bigg \lfloor \frac{\pi}{4 \arcsin(\sqrt[]{ 0.0124})}\bigg \rfloor  = 7,
\end{align}
and

\begin{align}
Q^7(\GoodBitVector = 00)\ket{\Psi} \approx \ket{E_{es}} \ket{E^{(1)}_{es}E^{(2)}_{es}\hdots E^{(20)}_{es} },
\end{align}
which was confirmed by the numerical experiment in figure \ref{fig. H2_PHILTER}A. After further iterations the probability drops and peaks at 21, 35, 49 etc, as expected. An interesting special case occurs when $\AnsatzProbOverlap = 1/4$. Of course, standard QPE can solve this problem efficiently, with high probability, but using the amplitude amplification method a solution is found with certainty after a single iteration. Here $\sin^2(\theta_\AnsatzProbOverlap) = 1/4 $ and therefore $\theta_\AnsatzProbOverlap = \pi/6$. It follows that $\cos(3 \theta_\AnsatzProbOverlap) = 0$ (equation (\ref{Eq. MatrixNotation})), which was confirmed by the numerical experiment in figure \ref{fig. H2_PHILTER}B. Another special case is if $\AnsatzProbOverlap = 1/2$, where the amplitude amplification method will neither increase nor decrease the amplitudes - each iteration rotates the state to its original state, which was confirmed by the numerical experiment in figure \ref{fig. H2_PHILTER}C. That is, for $\AnsatzProbOverlap> 1/2$ the good states cannot be amplified, as expected.  Figure \ref{fig. H2_PHILTER}D shows the spread in QPE amplitudes for the ground and excited state energy for increasing number of qubits in the energy register.  The  numerical calculations show for $m = 2$, a large portion of the unwanted ground state energy is associated with the target interval. Increasing the number of qubits in the energy register narrow down the width, and for $m = 10$, we observe a sharp peak at the energy eigenvalues, allowing almost only the correct bit-string to be measured. For successful amplification, i.e. amplifying the target excited state,  we need at least 6 qubits in the energy register, as shown in figure \ref{fig. H2_PHILTER}E. The plot shows which of the two energy eigenstates encompass more of the probability associated with the bit-string $\GoodBitVector =00$ in the energy register after running the QPE. The target excited state shows an almost constant probability at the initial success probability, as expected, since the sum of QPE amplitudes approximate to one, $\sum_{\QPEbitstring_i\in X_{00}} | \epsilon_{\QPEbitstring_i}^{(es)}|^2\approx 1$.  The probability associated with the ground state decreases with increasing number of qubits, as expected, by suppression of the QPE amplitudes. That is, for a total of six qubits in the energy register, we have more of the bit-string $\GoodBitVector =00$ associated with the target excited state compared to the ground state, thus amplifying the bit-string $\GoodBitVector =00$ results in successful amplification.  We refer the reader to \ref{app. example} for more details about figure  \ref{fig. H2_PHILTER}E. Consequently, the iterative PHILTER protocol (section \ref{sec:iaa}) needs at least six qubits to amplify the target excited state, as shown in figure \ref{fig. H2_PHILTER}F. According to equation (\ref{eq.iaa-qubits}), we would need $\stabilizer =  \lceil \log_2 \big( 2/0.0124\big)  - 2+ \PHILTERtolerance   \rceil =\lceil 5.33 + \PHILTERtolerance \rceil  $ plus the additional two qubits $\LenGoodBitVector =2$ in the energy register for successful amplification. The overestimation of the number of qubits by equation (\ref{eq.iaa-qubits}) is due to formula being an upper bound (\ref{app.stabilizer-qubits}). 
\begin{figure}[t]  
\centering \includegraphics[width=1.0\textwidth]{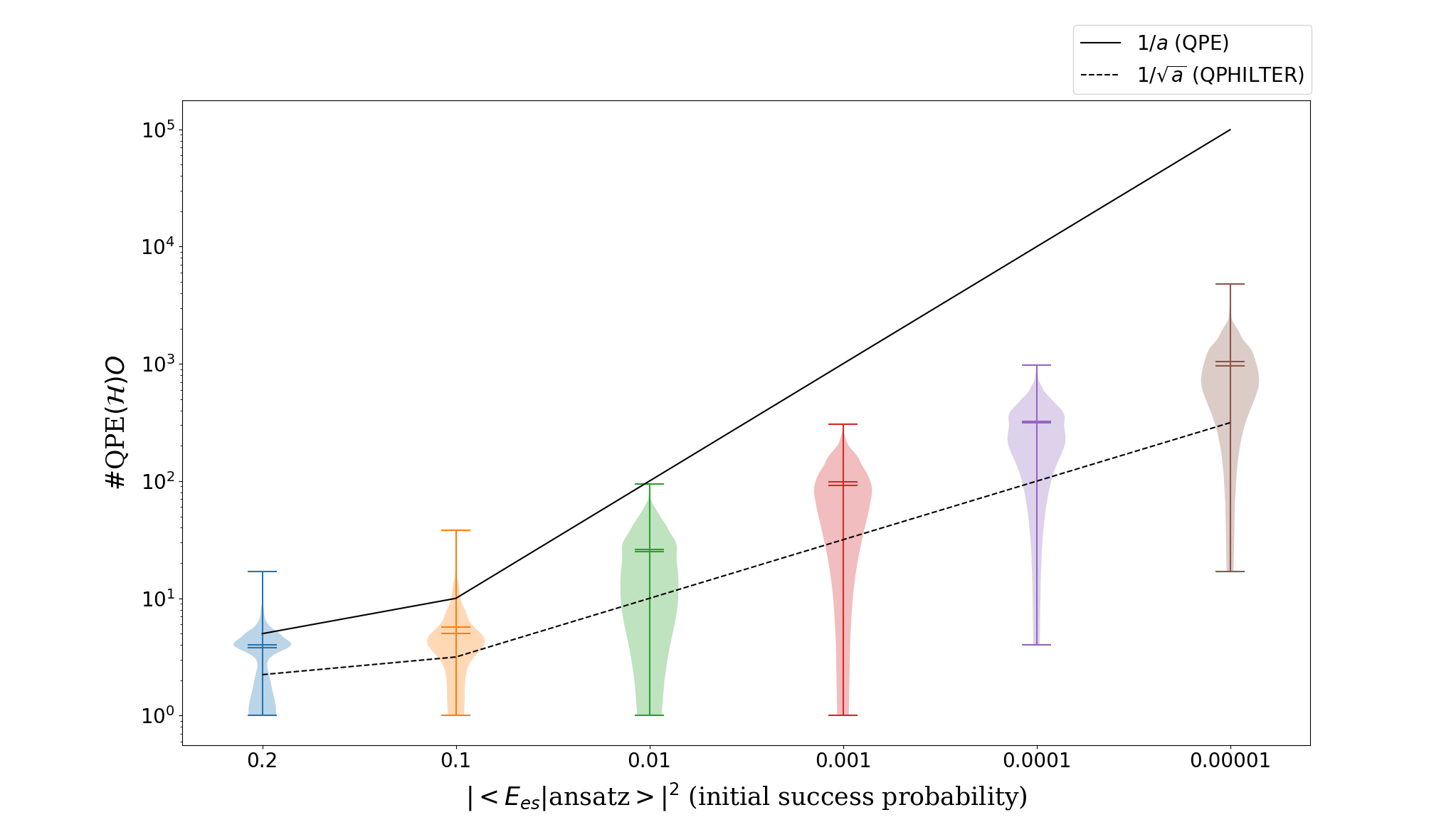} 
 \caption{Violin plots illustrating the probability distributions of the runtime obtained from the \textsf{Q}PHILTER protocol for molecular Hydrogen in two spatial orbitals (STO-3G). The runtime is in units of $\text{QPE}(\mathcal{H})O$ for measuring an energy with $E \in \good$.  Each violin plot consists of 1000 runs of the \textsf{Q}PHILTER protocol. The distance between 0.2 and 0.1 is inconsistently scaled to prevent the two violin plots to overlap. }
\label{fig. H2_QPHILTER}
\end{figure}
Figure \ref{fig. H2_QPHILTER} shows numerical data for the runtime of the \textsf{Q}PHILTER  protocol (algorithm \ref{AlgQsearch}) as a function of overlap with the target excited state. We compare the numerical data with the scaling of QPE and \textsf{Q}PHILTER. That is, using standard QPE, we shall expect to repeat state preparation with $\text{QPE}(\mathcal{H})O$ $O(1/\AnsatzProbOverlap)$ times on average before a state with $E \in \good$ is found. We assume it is not possible to run in parallel when comparing with  \textsf{Q}PHILTER. The expected runtime for \textsf{Q}PHILTER is  $O(1/\sqrt{\AnsatzProbOverlap})$ in units of $\text{QPE}(\mathcal{H})O$. We test the  \textsf{Q}PHILTER protocol by constructing ans\"{a}tze with decreasing overlap with the target excited state. Clearly for $\AnsatzProbOverlap = 10^{-5}$, the  \textsf{Q}PHILTER method shows its power over standard QPE by finding the excited state energy in order of magnitude less in runtime.  For increasing overlap between the ground and excited state, we find the gap between standard QPE and \textsf{Q}PHILTER is shrinking, especially for $\AnsatzProbOverlap =  0.2$, where the runtime is similar in both cases, however, the circuit depth in \textsf{Q}PHILTER is much greater than standard QPE. 

\section{Discussion}

\begin{figure}[t]  
\centering
\includegraphics[width=0.9\textwidth]{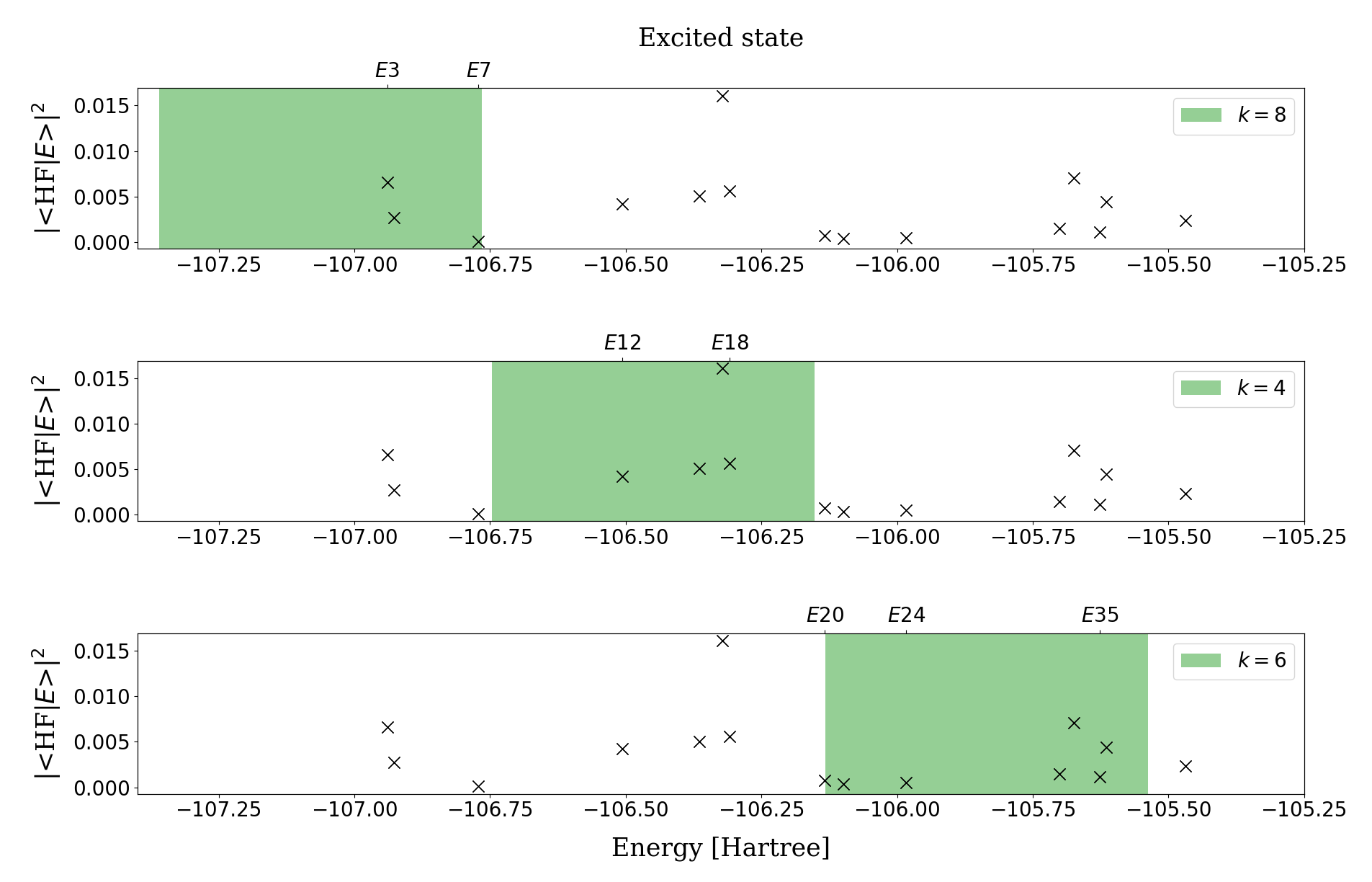} 
\caption{ Molecular nitrogen ($\text{N}_2$) in STO-3G basis. We use the Psi4 package\cite{turney2012,parrish2017} to obtain the overlap between the Hartree-Fock (HF) state and the energy eigenstates  at equilibrium bond length 1.098$\text{\r{A}}$. The top axis labels some of the energy eigenstates, for example $E7$ corresponds to the 7th excited state in the 14 electron singlet subspaces.  The green area denotes the target energy-interval and the corresponding \emph{k}-value, given in equation (\ref{Eq.general_ iterations}), gives the number of iterations of $Q(\mathcal{A}, \GoodBitVector) $ needed to obtain any energy eigenvalue within the target interval with almost unity probability. The target intervals are $\GoodBitVector =  0010101110$ (top), $\GoodBitVector =  0010101101$ (middle) and  $\GoodBitVector =  0010101100$ (bottom), and $\LenGoodBitVector = 10$. The target intervals were obtained by scaling the energy eigenvalues with $1/100$ to be between $[0, 2\pi)$.    }
\label{fig. N2}
\end{figure}

In our demonstration (section \ref{sec. numerical}) the algorithm uses up to 20 qubits for the molecular hydrogen system which could be solved within a variational quantum eigensolver (VQE) algorithm using only 1 to 4 qubits, depending on the applied qubit encoding. In this work we actually used a compressed encoding for the state register, so that the actual molecular wave function is sufficiently represented by a single qubit. The other qubits are needed for the numerical representation of the energy in QPE. The size of the energy register will however remain constant for larger molecules and the overall qubit requirements will scale comparably between VQE and QPE based treatments. One key advantage of QPE is that it allows sampling of \emph{m}-bit approximations of exact eigenvalues from initial trial states that can be far from the target eigenstate, while VQEs aim to explicitly prepare states. Especially for high lying excited states this is challenging for VQEs since most algorithms require sequentially solving the whole spectrum up to the desired energy where each lower lying state has to be represented with high accuracy\cite{lee2019}. For example, consider molecular nitrogen in figure \ref{fig. N2}. If the target state is the 35th excited state ($E35$), then we need to solve for all the 34 states below this target state with the sequential VQE. In addition, the lower lying states need to be solved very accurately (otherwise the projector becomes ill defined and errors might accumulate with every new state in the sequential procedure), which is the reason why this method is expected to perform well for low lying excited states. Using the PHILTER protocol, we would apply $Q^6(\mathcal{A}, \GoodBitVector)$ on the initial state, where we drive the HF state to the target energy-interval (containing $E35$), and measure the energies within that energy-interval with almost unity probability.

\section{Conclusion}

In this article, we have proposed algorithms for a quantum computer to discover the spectra of Hamiltonians by sampling the set of energies within a target energy-interval. An advantage of our method is that we do not require good approximations of the target energy eigenstates. Thus these algorithms are designed for cases where good approximations for the target energy-interval are either unknown or hard to prepare. We tested our method on the electronic Hamiltonian of molecular Hydrogen in a minimal representation (STO-3G), and successfully obtained the excited state energy given ans\"{a}tze with small overlap with the excited state. Our method is not limited to molecular Hamiltonians $-$ the algorithm can be applied to any Hamiltonian with the purpose of determining energy eigenvalues within a target energy-interval. In future work, we will explore  how the recent algorithms in amplitude estimation\cite{grinko2019, suzuki2020,  wie2019, aaronson2020} interplay with the algorithm proposed in this paper. 

\section*{Acknowledgement}
We thank Abhinav Anand, Sumner Alperin and Patrick Rall for helpful discussions. Al\'{a}n Aspuru-Guzik and his research group acknowledge the generous support from Google, Inc. in the form of a Google Focused Award. Al\'{a}n Aspuru-Guzik acknowledges the Vannevar Bush Faculty Fellowship under contract ONR N00014-16-1-2008. This research was supported by the Carl og Ellen Hertz's Legat til dansk l\ae ge- og naturvidenskab, Augustinus fonden, Henry og Mary Skovs Fond, Knud H\o jgaards Fond, Viet-Jacobsens fonden and U.S. Department of Energy through grant $\#$DE-AC02-05CH11231 subgrant LBNL - $\#$505736. The Quantikz package was used for typesetting quantum circuit diagrams with \LaTeX\cite{kay2018}. Computations were performed on the Niagara supercomputer at the SciNet HPC Consortium. SciNet is funded by: the Canada Foundation for Innovation; the Government of Ontario; Ontario Research Fund - Research Excellence; and the University of Toronto\cite{Loken2010, Ponce2019}. We thank the generous support of Anders G. Fr\o seth.

\appendix

\section{Grover's search algorithm}
\label{App:Grovers_search_algorithm}
The amplitude amplification process is a generalization of the Grover's search algorithm\cite{Grover1996, Grover1997}. Consider the case $\mathcal{A} = H$, where \emph{H} is the Hadamard transform acting in parallel on \emph{l} qubits. Then $(H\ket{0})^{\otimes l}= \frac{1}{2^{l/2}} \sum_{\QPEbitstring = 0}^{2^l -1} \ket{\QPEbitstring}$, where $\QPEbitstring$ is an integer $\QPEbitstring = \{0,\hdots,2^l -1\}$ and $\ket{\QPEbitstring}$ denotes the binary representation of $\QPEbitstring$. This is an equally weighted superposition of all $2^l$ states written in the computational basis states $\{ \ket{\QPEbitstring} \}$. If one of these computational basis states is a solution to our search problem, then our chances of classically guessing the right state $\ket{\QPEbitstring^*}$ is 1 in $2^l$. Thus the classical complexity is $O(2^l)$, where the quantum complexity turns out to be $O(\sqrt[]{2^l})$. The operator $Q(\GoodBitVectorgeneral)$ is equal to the iterate 

\begin{align}
Q(\GoodBitVectorgeneral) = -H S_0 H S_\GoodBitVectorgeneral \quad \text{(Grover iteration)}
\end{align}
where $S_\GoodBitVectorgeneral$ recognizes the solution $\ket{\QPEbitstring^*}$. This is the Grover's original search algorithm. Note the Hadamard gate is its own inverse.

\section{Qubit requirement for successful amplification}
\label{app.stabilizer-qubits}
Let $\ket{\Psi} = \text{QPE}\big(\mathcal{H}\big)\emph{O}\ket{0}^{\otimes (m+n)}$, where \emph{m} is the number of qubits in the energy register for QPE. The probability of observing the computational basis state $\ket{\QPEbitstring_i}$ in the \emph{m}-qubit register is then given by the expectation value of $M_{\QPEbitstring_i} =  \ket{\QPEbitstring_i}\bra{\QPEbitstring_i}$, where $\QPEbitstring_i$ is an integer $\QPEbitstring_i \in \{0, \hdots, 2^m-1 \}$ and $\ket{\QPEbitstring_i}$ denotes the binary representation of $\QPEbitstring_i$, as

\begin{align}
P(\Delta_{\tilde{E}_j, \QPEbitstring_i}) = \bra{\Psi} (I \otimes M_{\QPEbitstring_i}) \ket{\Psi} =  \sum_{j} |a_j|^2 \frac{\sin^2(\pi 2^m \Delta_{\tilde{E}_j, \QPEbitstring_i})}{2^{2m} \sin^2(\pi \Delta_{\tilde{E}_j, \QPEbitstring_i})},
\end{align}
where $\Delta_{\tilde{E}_j, \QPEbitstring_i} \equiv (\tilde{E}_j -\QPEbitstring_ i + \delta)/2^m$ and $\tilde{E}_j$ is an integer $\tilde{E}_j\in \{0,\hdots,2^{m}-1 \}$ such that $\tilde{E}_j / 2^m$ is the best \emph{m}-bit approximation to the energy eigenvalue $E_j /2^m$ within an accuracy of $2^{-m}$, that is

\begin{align}
\frac{E_j}{2^m}= \frac{\tilde{E}_j}{2^m}+ \frac{\delta}{2^m},
\end{align}
where $0  \leq \delta < 1$. The QPE amplitudes are therefore

\begin{align}
|\epsilon_{\QPEbitstring_i}^{(j)}|^2 = \frac{\sin^2(\pi 2^m \Delta_{\tilde{E}_j, \QPEbitstring_i})}{2^{2m} \sin^2(\pi \Delta_{\tilde{E}_j, \QPEbitstring_i})}. \label{app.sq-amplitude}
\end{align}
For the amplification process to be successful, i.e. to reduce unwanted energy-eigenstates, the following condition must be met

\begin{align}
\sum_{\substack{E_j \in \good, \QPEbitstring_i \in \GoodSet}} |a_j \epsilon^{(j)}_{\QPEbitstring_i}|^2  > \sum_{\substack{E_j \notin \good, \QPEbitstring_i \in \GoodSet}} |a_j \epsilon^{(j)}_{\QPEbitstring_i}|^2. \label{app.eq-old-condition}
\end{align}
We denote the bit-string that we amplify by $\GoodBitVector$. Consider $E_j \notin \good$ and $\QPEbitstring_i \in \GoodSet $, i.e. the probability of unwanted energy eigenstates to collapse to computational basis-states starting with the correct bit-string $\GoodBitVector$. The probability to getting an error  $|\tilde{E}-\QPEbitstring_i | \geq \QPEerrorTolerance $ for $\QPEerrorTolerance \in \mathbb{Z}^+/\{1\} $ and $\QPEbitstring_i\in \GoodSet$ is

\begin{align}
&\sum_{E_j \notin \good } |a_j|^2  \bigg(\sum_{\{\QPEbitstring_i \in \GoodSet: \hspace{0.1cm}\QPEerrorTolerance \leq (\tilde{E}_j-\QPEbitstring_i) < 2^{m-1}\}} |\epsilon^{(j)}_{\QPEbitstring_i}|^2  +\sum_{\{\QPEbitstring_i \in \GoodSet: \hspace{0.1cm}-2^{m-1} \leq (\tilde{E}_j-\QPEbitstring_i) < -\QPEerrorTolerance\}} |\epsilon^{(j)}_{\QPEbitstring_i}|^2 \bigg)  \label{app.eq.step1} \\
< &\sum_{E_j \notin \good } |a_j|^2  \frac{1}{2(\QPEerrorTolerance-1)} \label{app.eq.step2} \\ 
< & \frac{1}{2(\QPEerrorTolerance-1)}.
\end{align}
We refer the reader to Appendix C in \cite{cleve1998} for a detailed analysis of going from step (\ref{app.eq.step1}) to (\ref{app.eq.step2}).  The condition (\ref{app.eq-old-condition}) is simplified to 

\begin{align}
\sum_{\substack{E_j \in \good, \QPEbitstring_i \in \GoodSet}} |a_j \epsilon^{(j)}_{\QPEbitstring_i}|^2 > \frac{1}{2(\QPEerrorTolerance(\mathcal{H}) -1)}. \label{app.eq. simplifycon}
\end{align}
The Hamiltonian-dependent value $\QPEerrorTolerance(\mathcal{H})$ is the smallest distance between the target energy-interval and an energy eigenvalue outside the target energy-interval. The distance  $\QPEerrorTolerance(\mathcal{H})$ increases exponentially with increasing precision of the measured energy, i.e. by increasing the number of qubits in the energy register for QPE.  For example, consider $\tilde{E}_j = 0100101\hdots$ and $\GoodBitVector = 00$ ($\good \in [0,1/4)$), that is $\tilde{E}_j$ is not within the target energy-interval, $E_j \notin \good$. The smallest distance between $\tilde{E}_j$ and $\good$ is 

\begin{align}
01 - 00 &= 1 \quad \text{(2 bits precision)} \nonumber \\
010 - 001 &= 1  \quad \text{(3 bits precision)}  \nonumber \\
0100 - 0011 &= 1  \quad \text{(4 bits precision)}  \nonumber \\
01001 - 00111 &= 2  \quad \text{(5 bits precision)}  \nonumber \\
010010 - 001111 &= 3   \quad \text{(6 bits precision)}  \nonumber \\
0100101 - 0011111 &= 6  \quad  \text{(7 bits precision)}  \nonumber \\
\vdots \nonumber 
\end{align}
where the left and right side is written in base 2 and 10, respectively. In practice, we cannot determine $\QPEerrorTolerance(\mathcal{H})$ since it depends on the spectrum of the Hamiltonian. However, if we agree to accept also energies sufficiently close to the target interval, i.e. energies within a larger interval $\good'>\good$, the required number of qubits can be bounded. Specifically, we define the acceptance interval  $\good'$ as the interval $\good$ expanded a distance $2^{-\PHILTERtolerance}$ to either side, where $1\leq \PHILTERtolerance $ with $\PHILTERtolerance \in \mathbb{R}$ is a parameter which determines our tolerance for error, as depicted in figure \ref{fig. DE}. Let $\QPEerrorTolerance(\mathcal{H}) = 2^{m-\PHILTERtolerance}$ and the condition  (\ref{app.eq. simplifycon}) is

\begin{align}
&\sum_{\substack{E_j \in \good, \QPEbitstring_i \in \GoodSet}} |a_j \epsilon^{(j)}_{\QPEbitstring_i}|^2>  \frac{1}{2^{m-\PHILTERtolerance+1} -2}.
\end{align}
Adding $\sum_{\substack{E_j \notin \good, \QPEbitstring_i \in \GoodSet}} |a_j \epsilon^{(j)}_{\QPEbitstring_i}|^2$ to each side of the above equation and using the definition in equation (\ref{eq. true prob}), we obtain

\begin{align}
&b>  \frac{1}{2^{m-\PHILTERtolerance} -1} \quad \rightarrow \quad m >  \log_2 \big( 1/\TrueProbOverlap + 1 \big) + \PHILTERtolerance.
\end{align}
Using $ \log_2( 1/\TrueProbOverlap+ 1 ) \leq  \log_2 ( 2/\TrueProbOverlap) $ since $\TrueProbOverlap \in [0,1]$, and letting $m =\LenGoodBitVector + \stabilizer$, where $\LenGoodBitVector$ is the number of bits we amplify and

\begin{align}
 \stabilizer = \bigg \lceil \log_2 \big( 2/\TrueProbOverlap\big)  + \PHILTERtolerance- \LenGoodBitVector  \bigg \rceil, \label{eq.app.mqubits}
\end{align}
the amplification protocol may not filter out all energy eigenvalues outside the target interval, but it is guaranteed to filter out any energy eigenvalues outside the larger interval $ \good'$ expanded by $2^{-\PHILTERtolerance}$ to either side. 

\section{Determination of the number of qubits for successful amplification  for molecular hydrogen}
\label{app. example}
The purpose of this section is to determine the number of qubits needed for energy register (the \emph{m}-qubit register) for the quantum phase estimation (QPE) in order for the amplification process to be successful, i.e. reduce the probability for unwanted states. In the specific case of molecular Hydrogen in the minimal basis STO-3G, the ground and excited state are given in (\ref{eq. H1}) and we will consider the Hartree-Fock ansatz,

\begin{align}
O\ket{0}^{\otimes n} = \ket{\text{HF}} = a_{gs} \ket{E_{gs}} + a_{es} \ket{E_{es}},
\end{align}
where $a_{gs} = 0.99377$ and $a_{es} = 0.11149$. The goal is to amplify the excited state. In order to extract information about the energies of the ground and excited state, assume we add another \emph{m}-qubit register and run the QPE algorithm. The result of this will be that the \emph{m}-qubit register stores a binary representation of a phase related to the energies of the state 

\begin{align}
\ket{\Psi}=&\text{QPE}(\mathcal{H}^{(1)}) O\ket{0}^{\otimes  (m+ n)}   = a_{gs} \ket{E_{gs}}  \bigg(\sum_{\QPEbitstring_i \in \{0,1\}^m}^{2^m} \epsilon^{(gs)}_{\QPEbitstring_i} \ket{\QPEbitstring_i} \bigg) + a_{es} \ket{E_{es}} \bigg( \sum_{\QPEbitstring_i \in \{0,1\}^m}^{2^m} \epsilon^{(es)}_{\QPEbitstring_i} \ket{\QPEbitstring_i} \bigg),  \label{AppStaEq1}
\end{align}
where $\epsilon^{(j)}_{\QPEbitstring_i}$ are complex amplitudes given in equation (\ref{app.sq-amplitude}). The energy eigenvalues of the energy-eigenstates in (\ref{eq. H1}), to a precision of 20 bits, are 

\begin{align}
E_{gs} &= 01001011101011100010    \quad \text{and} \quad E_{es} = 00001001001001101111,
\end{align}
where the rescaling factor $-2 \pi E_h/ 2^{20}$  converts the bit-strings to energies. The bits are numbered from left to right starting from the most significant bit. The two energies differ for the second most significant bit. Thus amplifying the two most significant bits $\GoodBitVector = 00$ would amplify the amplitude associated with the excited state. 
\hfill \break
\hfill \break
In the following, we will compute and compare the QPE amplitudes for the two energy eigenstates, starting with two qubits in the energy register of the QPE, as well as compare the results with equation (\ref{eq.app.mqubits}). 
\hfill \break
\hfill \break
$\boldsymbol{ m = 2}$: In the case of using two qubits in the energy register for QPE ($m = 2$), then the best 2-bit approximation to the ground state is $\tilde{E}_{gs} = 01$. The remainder is given by $\delta_{gs} = 2^2 (309986/2^{20} - 1/2^{2}) = 0.1825$, where $E_{gs}$ is  converted  to base 10, i.e. $E_{gs} = 309986$. The probability of observing the computational basis state $\ket{\QPEbitstring_i}$ in the 2-qubit register is then given by the expectation value of $M_{\QPEbitstring_i} = \ket{\QPEbitstring_i}\bra{\QPEbitstring_i}$, where $\QPEbitstring_i$ is an integer $\QPEbitstring_i\in \{0,1,2,3\}$ and $\ket{\QPEbitstring_i}$ denotes the binary representation of $\QPEbitstring_i$. The QPE amplitude probabilities are 

\begin{equation}
|\epsilon^{(gs)}_{\QPEbitstring_i}|^2= \frac{\sin^2(\pi 2^2 \Delta_{\QPEbitstring_i,\tilde{E}_{gs}})}{2^{2\cdot 2} \sin^2(\pi \Delta_{\QPEbitstring_i,\tilde{E}_{gs}})} = 
    \begin{cases}
      0.029 & \text{for $\QPEbitstring_i$ = 00}\\
      0.90 & \text{for $\QPEbitstring_i$ = 01}\\
      0.051 & \text{for $\QPEbitstring_i$ = 10} \\
      0.019 & \text{for $\QPEbitstring_i$ = 11} ,
    \end{cases}       
\end{equation}
where $\Delta_{\QPEbitstring_i,\tilde{E}_{gs}} = (\tilde{E}_{gs}- \QPEbitstring_i + 0.1825)/4$ and $\tilde{E}_{gs} = 01$ is the 2-bit approximation to the ground state energy. Including the factor $|a_{gs}|^2 = |0.99377|^2$ from the ansatz, we obtain the final probabilities associated with the ground state,

\begin{equation}
|0.99377\cdot\epsilon^{(gs)}_{\QPEbitstring_i}|^2  =
    \begin{cases}
     0.029 & \text{for $\QPEbitstring_i$ = 00}\\
    0.88 & \text{for $\QPEbitstring_i$ = 01}\\
     0.050 & \text{for $\QPEbitstring_i$ = 10} \\
     0.019 & \text{for $\QPEbitstring_i$ = 11}.
    \end{cases}       
\end{equation}
Thus the probability to obtain the best 2-bit approximation to the ground state is 0.88. The best 2-bit approximation to the excited state is $\tilde{E}_{es} = 00$. The remainder is given by $\delta_{es} = 2^2 (37487/2^{20} - 0/2^{2}) = 0.1430$, where $E_{es} = 37487$ in base 10. Thus the QPE probabilities are

\begin{equation}
|\epsilon^{(es)}_{\QPEbitstring_i}|^2 = \frac{\sin^2(\pi 2^2 \Delta_{\QPEbitstring_i,\tilde{E}_{es}})}{2^{2\cdot 2} \sin^2(\pi \Delta_{\QPEbitstring_i,\tilde{E}_{es}})} =
    \begin{cases}
      0.94 & \text{for $\QPEbitstring_i$ = 00}\\
      0.030 & \text{for $\QPEbitstring_i$ = 01}\\
      0.012 & \text{for $\QPEbitstring_i$ = 10} \\
      0.020 & \text{for $\QPEbitstring_i$ = 11} 
    \end{cases}       
\end{equation}
where $\Delta_{\QPEbitstring_i,\tilde{E}_{es}} = (\tilde{E}_{es} - \QPEbitstring_i+ 0.1430)/2^2$. Including the factor $|a_{es}|^2 = |0.11149|^2$, we obtain the final probabilities associated with the excited state,

\begin{equation}
|0.11149 \cdot\epsilon^{(es)}_{\QPEbitstring_i}|^2  =
    \begin{cases}
     0.012 & \text{for $\QPEbitstring_i$ = 00}\\
    3.7\cdot 10^{-4}  &  \text{for $\QPEbitstring_i$ = 01}\\
    6.2\cdot 10^{-4}  & \text{for $\QPEbitstring_i$ = 10} \\
     2.4\cdot 10^{-4} & \text{for $\QPEbitstring_i$ = 11}. 
    \end{cases}       
\end{equation}
The important point is that if we were to amplify the bit string $\GoodBitVector = 00$ in order to amplify the excited state and reduce the ground state using 2 qubits in the energy register for the QPE, then the amplification process would be unsuccessful. The reason is that the probability would still be greater to obtain the ground state because

\begin{align}
\underbrace{|a_{gs}\cdot\epsilon^{(gs)}_{00}|^2}_{= 0.029} > \underbrace{|a_{es}\cdot\epsilon^{(es)}_{00}|^2}_{=0.012}.
\end{align}
That is, we have ``more'' of the bit-string 00 associated with the ground state compared to the excited state due to the small amplitude $a_{es}$. In conclusion, two qubits ($m = 2$) is not enough to reduce the probability for the ground state hence, unsuccessful amplification occurs. 
\hfill \break
\hfill \break
$\boldsymbol{ m = 3}$:  The best 3-bit approximation to the ground state is $\tilde{E}_{gs} = 010$. The remainder is given by $\delta_{gs} = 2^3 (309986/2^{20} - 2/2^{3}) = 0.3650 $, and the sum of the \emph{important} QPE probabilities are

\begin{align}
|0.99377|^2 \sum_{p = 0,1} |\epsilon^{(gs)}_{00p}|^2 = 0.070.
\end{align}
The best 3-bit approximation to the excited state is $\tilde{E}_{es} = 000$. The remainder is given by $\delta_{es} = 2^3 (37487/2^{20} - 0/2^{3}) = 0.2860$, and we find

\begin{align}
|0.11149|^2 \sum_{p = 0,1} |\epsilon^{(es)}_{00p}|^2 = 0.011.
\end{align}
Still, we have more of the bit-string $\GoodBitVector= 00$ associated with the ground state compared to the excited state, hence using 3 qubits in the first register is not enough to reduce the probability of the ground state, hence unsuccessful amplification occurs. 
\hfill \break
\hfill \break
In general, to have successful amplification the following condition must be met

\begin{align}
|0.99377|^2 \sum_{p,q\hdots =0,1} |\epsilon^{(gs)}_{00pq\hdots}|^2 < |0.11149|^2  \sum_{p,q\hdots = 0,1} |\epsilon^{(es)}_{00pq\hdots}|^2. \label{app-dsq-con}
\end{align}
Tables  \ref{table:app} and  \ref{table:app1}  show values up to $m = 12$. At $m = 6$, the condition in (\ref{app-dsq-con}) is met and we therefore need at least 6 qubits for successful amplification. According to equation (\ref{eq.app.mqubits}), the result is

\begin{align}
\stabilizer = \bigg\lceil \log_2 \bigg( \frac{2}{0.0124} \bigg) +\PHILTERtolerance-2 \bigg\rceil = \bigg\lceil 5.33 +\PHILTERtolerance \bigg\rceil  \label{eq.app.h2-stabili}
\end{align}
plus the additional 2 qubits $(\LenGoodBitVector= 2)$. Equation (\ref{eq.app.mqubits}) clearly overestimates the number of qubits needed for successful amplification, as expected since the formula is an upper bound.

\newcommand{\red}[1]{{\color{red} #1}}
\newcommand{\green}[1]{{\color{ForestGreen} #1}}
\renewcommand{\arraystretch}{1.5}
\begin{table}[H]
\caption{\label{table:app} Data for the ground state of molecular Hydrogen in STO-3G basis. The first column is the number of qubits in the energy register for the quantum phase estimation (QPE). The second column shows the difference between the true ground state energy (to 20-bits precision) and the \emph{m}-bit approximation, with $0 \leq \delta< 1$. The third and fourth columns show the QPE probability of the ground state to output computational basis-states starting with the correct bit-string $\GoodBitVector = 00$, described by the set $X_{00}$, with $a_{gs} = 0.99377$.}
\begin{tabular*}{\textwidth}{@{}l*{12}{@{\extracolsep{2pt plus
10pt}}l}}
&&Ground state \\
\br
\emph{m}&$\delta_{gs}$&$\sum_{\QPEbitstring_i \in X_{00}}|\epsilon_{\QPEbitstring_i}^{(gs)}|^2$&$\sum_{\QPEbitstring_i \in X_{00}}|a_{gs} \cdot \epsilon_{ \QPEbitstring_i}^{(gs)}|^2$\\
\mr
2&0.18&0.029&0.029 \\
3&0.37&0.071&0.070 \\
4&0.73&0.036 & 0.036  \\
5&0.46&0.043 &0.042\\
6&0.92&$1.6 \cdot 10 ^{-3}$&$1.6 \cdot 10 ^{-3}$\\
7&0.84&$3.2 \cdot 10 ^{-3}$ & $3.2 \cdot 10 ^{-3}$\\
8&0.68&$5.2 \cdot 10 ^{-3}$& $5.2 \cdot 10 ^{-3}$\\
9&0.36&$3.1 \cdot 10 ^{-3}$& $3.0 \cdot 10 ^{-3}$\\
10&0.72&$1.1 \cdot 10 ^{-3}$ & $1.1 \cdot 10 ^{-3}$\\
11&0.44&$9.2 \cdot 10 ^{-4}$ & $9.1 \cdot 10 ^{-4}$\\
12&0.88&$6.5 \cdot 10 ^{-5}$&$6.4 \cdot 10 ^{-5}$\\
\br
\end{tabular*}
\end{table}

\begin{table}[H]
\caption{\label{table:app1} Data for the excited state of molecular Hydrogen in STO-3G basis. The first column is the number of qubits in the energy register for the quantum phase estimation (QPE). The second column shows the difference between the true excited state energy (to 20-bits precision) and the \emph{m}-bit approximation, with $0 \leq \delta < 1$. The third and fourth columns show the QPE probability of the excited state to output computational basis-states starting with the correct bit-string $\GoodBitVector = 00$,  described by the set $X_{00}$, with $a_{es} = 0.11149$. The colors in the first column highlight when the condition in  (\ref{app-dsq-con}) is met (green), by comparing with table \ref{table:app}.}
\begin{tabular*}{\textwidth}{@{}l*{12}{@{\extracolsep{2pt plus
10pt}}l}}
&&Excited state \\
\br
\emph{m}&$\delta_{es}$&$\sum_{\QPEbitstring_i \in X_{00}}|\epsilon_{\QPEbitstring_i}^{(es)}|^2$&$\sum_{\QPEbitstring \in X_{00}}|a_{es} \cdot \epsilon_{ \QPEbitstring_i}^{(es)}|^2$\\
\mr
\red{2}&0.14&0.94&0.012 \\
\red{3}&0.29&0.88&0.011\\
\red{4}&0.57&0.88 & 0.011 \\
\red{5}&0.14&0.97 &0.012 \\
\green{6}&0.29&0.97&0.012 \\
\green{7}&0.58&0.98  & 0.012 \\
\green{8}&0.15&0.99& 0.012 \\
\green{9}&0.30&0.99 & 0.012 \\
\green{10}&0.61&0.99  & 0.012 \\
\green{11}&0.22&0.99 & 0.012 \\
\green{12}&0.43&0.99 &0.012 \\
\br
\end{tabular*}
\end{table}

\clearpage

\section*{References}
\bibliographystyle{iopart-num}
\bibliography{library}
\end{document}